\RequirePackage{fix-cm}
\RequirePackage{rotating}

\documentclass[smallextended]{svjour3}       
\smartqed  

\usepackage{csquotes}
\usepackage{booktabs} 
\usepackage{framed}
\usepackage{graphicx}
\usepackage{amssymb}
\usepackage{tabularx}
\usepackage{ltablex}
\usepackage{booktabs}
\usepackage{graphicx}
\usepackage{url}
\usepackage{tabulary}
\usepackage{multirow}
\usepackage{enumitem}
\usepackage{hyperref}
\usepackage{hypcap}
\usepackage{lscape}
\usepackage[normalem]{ulem}
\useunder{\uline}{\ul}{}
\usepackage{longtable}
\usepackage{color}
\usepackage{hyperref}

\usepackage{booktabs}
\usepackage[flushleft]{threeparttable}
\usepackage{microtype}

\usepackage{placeins}

\usepackage{siunitx}  
\usepackage{textgreek}

\usepackage{flushend}

\usepackage{lscape}

\usepackage{rotating}

\begin{document}

\title{Satisfaction and Performance of Software Developers during Enforced Work from Home in the COVID-19 Pandemic}

\titlerunning{Satisfaction and Performance of Software Developers during COVID-19}

\author{Daniel Russo\protect\footnote{These authors contributed equally.}       \and
        Paul H. P. Hanel$^{\ast}$
        \and
        Seraphina Altnickel 
        \and
        Niels van Berkel 
}


\institute{D. Russo  \at
            Corresponding author.\\
              Aalborg University, Department of Computer Science, Copenhagen, Denmark \\
              Tel.: (+45) 9940 7765\\
              \email{daniel.russo@cs.aau.dk}           
           \and
           P. H. P. Hanel \at
              University of Essex, Department of Psychology, Colchester, UK
                         \and
           S. Altnickel \at
              Think distributed, Berlin, Germany
                         \and
                  N. van Berkel \at
              Aalborg University, Department of Computer Science, Aalborg, Denmark
}

\date{Received: DD Month YEAR / Accepted: DD Month YEAR}

\maketitle

\begin{abstract}
Following the onset of the COVID-19 pandemic and subsequent lockdowns, the daily lives of software engineers were heavily disrupted as they were abruptly forced to work remotely from home. To better understand and contrast typical working days in this new reality with work in pre-pandemic times, we conducted one exploratory ($N$ = 192) and one confirmatory study ($N$ = 290) with software engineers recruited remotely. Specifically, we build on self-determination theory to evaluate whether and how specific activities are associated with software engineers' satisfaction and productivity. To explore the subject domain, we first ran a two-wave longitudinal study. We found that the time software engineers spent on specific activities (e.g., coding, bugfixing, helping others) while working from home was similar to pre-pandemic times. Also, the amount of time developers spent on each activity was unrelated to their general well-being, perceived productivity, and other variables such as basic needs. 
Our confirmatory study found that activity-specific variables (e.g., how much autonomy software engineers had during coding) do predict activity satisfaction and productivity but not by activity-independent variables such as general resilience or a good work-life balance. 
Interestingly, we found that satisfaction and autonomy were significantly higher when software engineers were helping others and lower when they were bugfixing. 
Finally, we discuss implications for software engineers, management, and researchers. 
In particular, active company policies to support developers' need for autonomy, relatedness, and competence appear particularly effective in a WFH context.

\end{abstract}

\section{Introduction}
\label{sec:intro}


The COVID-19 pandemic has abruptly and unprecedentedly disrupted software developers' working routines.
On short notice, many software developers were asked to switch from their typical office-based working habits to a new working from home (WFH) setting.
This change in work setting has had a considerable negative impact on developers' well-being and productivity~\cite{Ralph2020pandemic}, as the pandemic and subsequent restrictions (e.g., lockdowns) restricted their basic needs, such as the need for autonomy, competence, or relatedness~\cite{cantarero2021affirming}.
Nevertheless, longitudinal research has also shown that software engineers can successfully adapt over time, suggesting that their well-being and productivity bounce back to pre-pandemic levels~\cite{ford2020tale,forsgren_2020,bao2020does,russo2020predictors,smite2021changes}.
This is encouraging, as 89\% of professionals would like to work from home at least one day per month after the pandemic~\cite{Walton2020NZadaptation}.
For this reason, major IT companies (e.g., Twitter, Microsoft, AirBnB, Uber, Facebook) informed their employees that they could work from home indefinitely (e.g., Twitter) or extended the remote work policies providing specific support (e.g., AirBnB)~\cite{BusinessInsider2020}.
Thus, research conducted during the pandemic will likely also be of value once current restrictions have been lifted.

Software professionals working remotely for an organization is not a new topic in software engineering.
In 1983, Olson defined remote work as an ``\textit{organizational work that is performed outside of the usual organizational confines of space and time}''~\cite{olson1983remote}.
This definition implies professionals' high degree of freedom with regards to scheduling their working hours, activities, and the location from which they work.
With the rise of the internet in the late 90s, scholars started researching the challenges and opportunities of remote work from home~\cite{pounder1998homeworking}. 
In these cases, professionals usually have a high degree of autonomy in terms of time but not in terms of space since they have chosen their homes as primary working spaces.
Generally speaking, researchers investigated specific software development practices, such as processes~\cite{guo2001special,deshpande2016remote} or communication~\cite{higa2000understanding} to better tailor remote work practices to business needs. 
Similarly, collaboration and characteristics of remote and asynchronous projects have been extensively studied by the Global Software Engineering community~\cite{herbsleb2007global,vsmite2010empirical}. 
Such studies typically focus on the interaction among software development teams co-located in different geographical areas.
However, the focus has been on software development teams working together on distributed projects. 
There is a growing agreement within the practitioners' community that working from home is different from working remotely on distributed projects~\cite{Aten2020WFH}. While working from home is understood as working from the primary address of residence, such as an apartment or house, working remotely is carried out typically in co-working spaces or in different settings where one lives.
The pandemic made many of us realize that some of the fears often associated with remote work (such as decreasing productivity) are often unfounded. 
Hence, anecdotal evidence driving top managerial decisions due to the lack of specific research~\cite{mesaglio_2020} should be supplemented with scholarly evidence.

So far, the authors of this paper have worked on a comprehensive research agenda to understand the effects of the COVID lockdown on software engineers. 
We started looking at the self-perceived well-being and productivity in the earlier months of the pandemic~\cite{russo2020predictors}.
Afterward, we tracked how a typical work day looks like, as also the distribution of work activities compared to pre-COVID times~\cite{russo2021daily}.
Eventually, we performed a two years long longitudinal study with six waves to assess the effects of the entire pandemic on software developers~\cite{russo2021understanding}.
Additionally, the first author also investigated software process-related changes while working from home~\cite{Cucolas2021Scrum}.


This paper studies whether professionals' needs influence their time on various activities. In their seminal paper, Ryan and Deci~\cite{ryan2000self} introduce self-determination theory, which describes the three innate psychological needs that motivate us and guide our behavior: need for autonomy, competence, and relatedness. The need for autonomy measures whether people feel independent; the need for competence whether people can complete various (challenging) activities; and need for relatedness assesses whether people feel appreciated by others close to them. 
Self-determination theory has frequently been used in the work context to predict job satisfaction and performance~\cite{gagne2005self}. 
For example, research established that all self-determination theory-related needs (need for autonomy, relatedness, and competence) positively correlate with job satisfaction and productivity~\cite{van2010capturing}.
By building on self-determination theory, we study how software engineers' activities changed during the pandemic using the activity taxonomy of Meyer et al.~\cite{meyer2019today}.

In line with other researchers who started to look at productivity of software engineers in a more holistic way~\cite{sadowski2019rethinking}, we are particularly interested in understanding whether specific activities contribute to their well-being and productivity in general and which factors contribute to their satisfaction and productivity while working on a particular task. For example, meetings can be resource-draining and be felt as burdensome by employees~\cite{allen2012meeting}. 
Furthermore, we also take social relations as an indicator of need for relatedness into account: People who feel that communication with their colleagues and line managers is of importance might be more inclined to spend time in meetings, helping others, and other social activities and report higher well-being because their need for relatedness is then more likely to be satisfied. 
Prior research which investigated predictors of well-being and stress in occupational settings~\cite{bhui2016perceptions,edwards2009value,mccalister2006hardiness} has not measured the specific activities that might have contributed to higher stress and lower levels of well-being. However, the type of activity someone is doing might contribute to higher stress levels beyond other factors identified by previous research, such as support by coworkers and supervisors~\cite{chyi2018prediction}. If we were to determine which specific activities are associated with higher or lower levels of stress or well-being, this would provide valuable information for future research investigating predictors of stress. 
We divided this study into an exploratory and a confirmatory part to investigate all these aspects. Both studies build on self-determination theory~\cite{ryan2000self}.

In the exploratory investigation, we first measured developers' activities and self-reported well-being and productivity to assess changes throughout the lockdown over a two-week period.
We compared wave 1 with wave 2 to assess our test-retest reliability and stability of the data captured.
In particular, we found that the time software engineers spent doing specific activities from home was comparable when working in the pre-pandemic office.
Nevertheless, we also reported significant mean differences, such as less time dedicated to meetings and breaks and more time spend on specification and documentation. 
Interestingly, the time people spent on each activity was unrelated to their general well-being, perceived productivity, and other variables.
In hindsight, this is not surprising because many factors affect our well-being and productivity. For example, well-being is impacted by a range of factors such as the quality of our relationships, personality, or situational factors (e.g., weather)~\cite{connolly2013some,diener2009science,russo2020predictors}, which makes it unlikely that spending an hour more or less on a specific activity will significantly impact well-being. 
However, what we believe is more likely to impact well-being and productivity, are activity-specific features, which is one of the primary motivations of the confirmatory study (i.e., what factors predict activity-specific well-being and productivity?). 

In the confirmatory study, we measured activity-specific well-being and productivity, as well as the activity-specific need for autonomy, competence, and relatedness (e.g., how productive professionals felt during the activity they spend the most time on a day). Additionally, we explored whether task-unrelated variables such as resilience or work-life balance act as moderators between activity-specific needs and activity-specific well-being and productivity (see below for a more detailed rationale). 
Our findings confirm the long-standing intuition that software engineers feel more autonomous while coding than while in meetings or writing emails.
Also, software engineers experience less satisfaction with bugfixing but helping others is a satisfaction booster.
We further characterized which activities resulted in higher feelings of satisfaction, productivity, autonomy, competence, and connectedness. Moreover, through combining both the exploratory and confirmatory study, methodological lessons can be learned: Only asking whether overall well-being and productivity, for example, are associated with time spent on specific activities, misses the impact different activities can have on people's well-being and productivity. Measuring activity-specific well-being and productivity levels overcomes this limitation.

In the remainder of this paper, we describe the related work in Section~\ref{sec:related}, followed by a description of our research design in Section~\ref{sec:design}. 
The analysis and related results of our analysis are described in Section~\ref{sec:analysis}.
Implications and recommendations for software engineers and organizations are outlined in Section~\ref{sec:discussion}.
Finally, we conclude this paper by presenting future research directions in Section~\ref{sec:conclusion}.

\section{Related Work}
\label{sec:related}

Research on behavioral and emotional aspects within the software engineering community is a relatively new but rising research topic~\cite{sanchez2019taking}.
Developers' behaviors and emotional states do play a substantial role in how they are going to perform their working activities~\cite{graziotin2015affect}.
For this reason, the community started to focus specifically on software engineers' behaviors~\cite{lenberg2015behavioral}, emotions~\cite{graziotin2014happy}, or personality traits~\cite{cruz2015forty,russo2020gender}.

Concerning the pandemic, there is widespread agreement that lockdowns have a negative influence on well-being~\cite{brooks2020,lunn2020using}.
Living in a lockdown during a pandemic has been linked to increased levels of anger, depression, emotional exhaustion, fear of infecting others or becoming infected, insomnia, irritability, loneliness, low mood, and post-traumatic stress disorders~\cite{sprang2013posttraumatic,hawryluck2004sars,lee2005experience,marjanovic2007relevance,reynolds2008understanding,bai2004survey,Tag2022PandemicEmotionRegulation}.
Furthermore, anxieties of infection~\cite{kim2015public,prati2011social}, a lack of supplies or not being treated~\cite{wilken2017knowledge}, and false or conflicting information~\cite{caleo2018factors} can all cause substantial stress and give rise to new approaches to regulate our emotions~\cite{Tag2022PandemicEmotionRegulation}. 
Furthermore, the psychological impacts of being quarantined may take years to manifest~\cite{brooks2020}.

Pre-COVID research, on the other hand, indicates that remote working is associated with improved work-life balance, creativity, productivity, reduced stress, and low carbon emissions due to the absence of commuting~\cite{owl_labs_2019,anderson2015impact,bloom2015does,vega2015within,baruch2000teleworking,cascio2000managing}.
However, there are several apparent downsides to remote work, like decreased teamwork and communication, loneliness, the sensation of always being 'online,' decreased motivation, and distractions at home~\cite{buffer2020,yang2021effects}.
Aside from such factors, estimates indicate that remote work will grow significantly in the next years~\cite{owl_labs_2019,gallup2020}. 

In the software engineering domain, several large software companies, such as Stack Overflow or Red Hat, have embraced working from home by designing \textit{ad hoc} schemes already before the start of the 2020-Corona pandemic~\cite{stackoverflow_2017,hat_2015}.
Organizations do so to increase their employees' job satisfaction and productivity while simultaneously reducing their operating expenses, such as office rent~\cite{felstead2017assessing,perez2002benefits}.
Several aspects of remote and distributed working have been (indirectly) investigated by the Global Software Engineering community well before the pandemic (e.g.,~\cite{vsmite2010empirical,herbsleb2001global,richardson2012process}).
To better frame this study theoretically, we looked into peer-reviewed publications in Scopus which explicitly focused on working from home (i.e., and not remote and distributed work). 
We made this choice to narrow down the subject matter and consider only articles whose primary focus is about working from home.
We identified thirteen relevant papers in total.
Considering the vast but recent impact of COVID-19, we also selected non-peer-reviewed pre-prints on arXiv.
Table~\ref{tab:SLR} summarizes prior studies of remote working issues related to software engineers.

\newcolumntype{x}{>{\hsize=.15\hsize}X}
\newcolumntype{y}{>{\hsize=.425\hsize}X}

\begin{table}[!htbp]
\centering
\caption{Overview of prior studies about software engineers working from home}
\label{tab:SLR}
\scriptsize
\begin{tabularx}{\textwidth}{@{}xyy@{}}
\toprule
\textbf{Study} & \textbf{Method} & \textbf{Findings} \\ \midrule
Russo et al. (2021)~\cite{russo2021understanding} & Sample study. 14-months 4-wave longitudinal study with 15 variables associated to developers' well-being and productivity. & Well-being increased over time during the lockdown and productivity remained stable. \\ 
Smite et al. (2021)~\cite{smite2021changes} & Sample study. Six corporate surveys conducted in four Scandinavian companies. & No significant change in productivity but WFH impacts some more then others. \\ 
Cucolaș \& Russo (2021)~\cite{Cucolas2021Scrum} & Multi-Methods study. Qualitative interviews and sample study of Scrum developers. After a theoretical model was induced from qualitative data, a sample study of 200 software engineers validated it with PLS-SEM. & Home-working environment is the most important variable for project success, and to improve WFH conditions, organizations should strengthen the need for developers' autonomy, competence, and relatedness. \\ 
Miller et al. (2021)~\cite{miller2021your} & Field study. Mixed-methods investigation of Microsoft developers. Two surveys collected information about working from home and team-related issues. Data were analyzed using different quantitative and qualitative techniques. & Communication and interaction with colleagues is a relevant predictor of developers' satisfaction and team productivity. \\ 
Butler \& Jaffe (2021)~\cite{butler2021challenges} & Field study. Diary study of 435 Microsoft developers over 10 weeks during the lockdown. Data were analyzed using different quantitative and qualitative techniques. & The largest identified challenges were meetings, overwork, and physical and mental health. On the other hand, participants appreciated having more family time and work flexibility. \\ 
Machado et al. (2021)~\cite{machado2021gendered} & Sample study. Mixed-methods investigation of 233 Brazilian software professionals. Data were analyzed using different quantitative and qualitative techniques. & The pandemic affected men and women differently. Organizations should accommodate women first when scheduling meetings. Organize uninterrupted work sessions and support childcare are also recommended. \\ 
Ford et al. (2020)~\cite{ford2020tale} & Field study. Mixed-methods investigation of 3,634 Microsoft developers. Two surveys collected qualitative and quantitative insights about WFH conditions during the COVID-19 lockdown. & Quality of family life and time improved; although WFH might have led to a lack of focus, poor work-life boundaries, communications, and sync issues, developers adapt over time. \\ 
Ralph et al. (2020)~\cite{Ralph2020pandemic} & Sample study. Large-scale cross-sectional study of 2,225 software developers globally working from home during the COVID-19 lockdown, surveying five variables. Data were analyzed using covariance-based structural equation modeling.   & Confirmation of a theoretical model. Professionals' well-being and productivity are suffering; well-being and productivity are strongly related to each other; women are disproportionately affected by this peculiar remote working setting. \\ 
Russo et al. (2020)~\cite{russo2020predictors} & Sample study. Longitudinal study involving 192 software engineers living in countries with comparable COVID-19 lockdown measures, surveying 51 variables. Data were analyzed using correlations, multiple linear regressions, and covariance-based structural equation modeling to assess predictive causal relations. & Well-being and productivity are related, professionals adapt to the condition over time, improving their well-being and productivity, introverts are disproportionately affected by the lockdown, no predictor variable was significantly able to causally explain the variance in well-being and productivity. \\ 
Ford et al. (2019)~\cite{ford2019remote} & Field study. Qualitative study interviewing three transgender software engineers to explore the interplay of gender identity and remote work. & Working from home enables the empowerment and identity disclosure of software professionals from marginalized communities. \\ 
James \& Griffiths (2014)~\cite{james2014secure} & Experimental simulation. Within an existing project, relevant working from home problems has been identified and addressed by developing and validating a specific solution. & Development of a mobile execution
environment to support a secure and portable working from home setting. \\ 
Guo (2001)~\cite{guo2001special} & Field study. Report of two qualitative surveys regarding software process improvement related to the distinctive characteristics of teleworking.  & Development of the \textit{Software Process Improvement approach for Teleworking Environment} (SPITE) model. Identification of 25 base practices to improve software processes when working from home. \\ 
Higa et al. (2000)~\cite{higa2000understanding} & Field study. Mixed-methods study at Fujitsu with 44 software engineers to investigate how the use of E-mail influences telework. To test the hypotheses, three hierarchical regression models were used. &  An effective use of E-mails by remote workers leads to better work distribution and work productivity. \\ 
Pounder (1998)~\cite{pounder1998homeworking} & Formal theory. Essay about security problems linked to telework. & This is the first paper that considers "homeworking" as a distinct working setting. It discusses the main security concerns and makes recommendations for organizations. \\ \bottomrule
\end{tabularx}
\end{table}

Most papers which focused on WFH were published in or after 2019 and are related to the COVID-19 pandemic.
From a methodological perspective, most studies have been field studies involving a single company (i.e., Fujitsu~\cite{higa2000understanding}, Baidu~\cite{bao2020does}, and Microsoft~\cite{ford2020tale,miller2021your,butler2021challenges,yang2021effects}).
Such real-world investigations aimed to understand the research phenomena by generating research hypotheses.
Three studies were conducted in a neutral setting on the opposite spectrum by asking participants a quantifiable judgment and analyzing such data through statistical techniques.
These six sample studies generalize their result on the entire software engineering population~\cite{Ralph2020pandemic,russo2020predictors,machado2021gendered,Cucolas2021Scrum,russo2021understanding,smite2021changes}.

Content wise, half of the papers are concerned with specific topics related to working from home, such as security~\cite{pounder1998homeworking,james2014secure}, process~\cite{guo2001special}, work productivity~\cite{higa2000understanding,lamarche2020socially}, and inclusion~\cite{ford2019remote}.
The other half mostly investigated well-being and productivity while working from home during the pandemic~\cite{ford2020tale,Ralph2020pandemic,russo2020predictors,butler2021challenges,machado2021gendered,russo2021understanding,smite2021changes} or productivity-related to projects' characteristics~\cite{bao2020does,Cucolas2021Scrum}.

Overall, the investigated topic is not new to the community.
However, from this short review, we noticed how scholars focused in particular on WFH topics due to the COVID-19 pandemic and the subsequent lockdown.
Indeed, future work is needed to support developers working in a lockdown environment or in a reality where pandemic waves are part of our everyday lives. 
Alternatively, more optimistically, software organizations will enforce hybrid work in a widely spread manner.
Therefore, we believe that this subject matter is of utter importance for software professionals' well-being and productivity in the years to come. 
This is also important because past research has shown that there are some mean differences between software engineers and the general population~\cite{russo2021anecdote}. In other words, we cannot assume that findings from other population types (e.g., employees at Microsoft, general population) generalize to software engineers.


\section{Research Design}
\label{sec:design}

Our design was guided by the relevant ACM SIGSOFT Empirical Standards for longitudinal and sample studies~\cite{ralph2020empirical}. 
First, we applied an exploratory longitudinal design already described in Russo et al.~\cite{russo2021daily}.
Subsequently, to overcome the methodological limitations of the exploratory study while gaining further insights into the associations of activities with activity-specific satisfaction, productivity, and basic needs, we employed a cross-sectional design. 

We formulate the following five main research questions which were guided by previous research and by self-determination theory~\cite{ryan2000self}:

\begin{quote}
    \textbf{Research Question 1}: Has the distribution of daily working activities of software engineers changed while WFH during the pandemic as compared to pre-pandemic daily working activities?
\end{quote}

\begin{quote}
    \textbf{Research Question 2}: Is the distribution of daily working activities related to well-being, productivity, and other variables?
\end{quote}

\begin{quote}
    \textbf{Research Question 3}: To what extent does Self-Determination Theory (i.e., the needs for autonomy, competence, and relatedness) predict software engineers' activity-specific satisfaction and productivity during the COVID-19 pandemic?
\end{quote}

\begin{quote}
    \textbf{Research Question 4}: To what extent are the associations between activity satisfaction and productivity moderated by resilience and company support during the COVID-19 pandemic? 
\end{quote}

\begin{quote}
    \textbf{Research Question 5}: Do software engineers' work activities while WFH during the pandemic affect their activity-specific well-being, productivity, and psychological needs?
\end{quote}

We designed the exploratory study to answer RQ1 and RQ2, whereas the confirmatory research was designed to answer RQ3 to RQ5.

Our first concern was to recruit software professionals for our exploratory study carefully.
To do so, we used a multistage selection process, detailed in Section \ref{SSec:Participants}.
We asked them to complete the same survey on two occasions.
Unique randomized IDs were assigned to participants to preserve their anonymity and match their responses from both waves. 
To address concerns about replicability and increase the reliability of our findings, we asked the same participants to complete all measures twice, two weeks apart. This allowed us to test whether the distribution of daily working activities has changed. At the same time, we asked participants to report how much time they spend on 15 activities and compared the responses with a pre-pandemic sample~\cite{meyer2019today}, which allowed us to test whether the distribution has changed since the onset of the first lockdown in 2020. To test RQ2 -- is the time spent on different activities correlated with well-being, productivity, and other variables -- we correlated the time spent on each activity with professionals' general well-being, productivity, and other variables.  

In a subsequent confirmatory study, we asked participants about their well-being, productivity, autonomy, competence, and relatedness to their co-workers while completing specific activities (e.g., ``how stressed were you while coding?"). Specifically, to test RQ3 -- whether the needs for autonomy, competence, and relatedness predict software engineers' activity-specific satisfaction and productivity -- we asked how satisfied, productive, autonomous, competent, and related with their co-workers' participants felt during working on a specific activity (e.g., coding). Our design allowed us to test RQ3 across all activities but also separately for each activity.   

Additionally, to investigate RQ4 -- whether the associations between autonomy, competence, and relatedness with activity satisfaction and productivity are moderated by resilience and company support -- we also included a range of conceptually related variables that measure facets of company support: caring leadership, work-life balance, empowerment, job enablement, soft company support, hard company support, and recognition. We expect that software engineers who are more resilient and receive higher company support are less likely to be affected by, for example, reduced autonomy for a specific task. For instance, resilience or recognition might buffer against reduced autonomy because resilient people are more likely to bounce back after stressful events such as being less able to make autonomous decisions~\cite{smith2008brief,weinstein2011self}. Further, software engineers who experience low autonomy, competence, or relatedness during their work will experience only lower satisfaction and be less productive if their company does not provide adequate support that helps to buffer against the negative impact. In other words, we expect the effect of the three needs on activity satisfaction and productivity to be reduced if resilience and company support is high. 

Finally, to test RQ5 -- does the activity impact activity-specific satisfaction, productivity, and psychological needs -- we tested during which activity professionals felt relatively more or less satisfied, productive, and so on. \par

\subsection{Theoretical Framework}

We are performing this investigation using the Self-Determination Theory (SDT) framework.
In particular, this theoretical framework has been used to design organizational policies to improve both well-being and high-quality performance~\cite{gagne2005self}.
SDT is a macro theory of human motivation that focuses, among others, on the motivations in the workplace~\cite{ryan2000self}.

The general idea of SDT is that if the three basic needs for competence, autonomy, and relatedness are satisfied, they lead to an increase in professionals' intrinsic motivation, productivity, and well-being. 
Indeed, employees' well-being is not only an ethical concern for every business but also a pivotal aspect to enhancing organizational sustainability, which is directly related to customers' satisfaction and financial success~\cite{mackey2014conscious}. 
As a macro theory, it includes several factors that lead to employees' well-being, such as the three basic needs. 

\begin{figure*}
  \centering
  \includegraphics[width=1\textwidth]{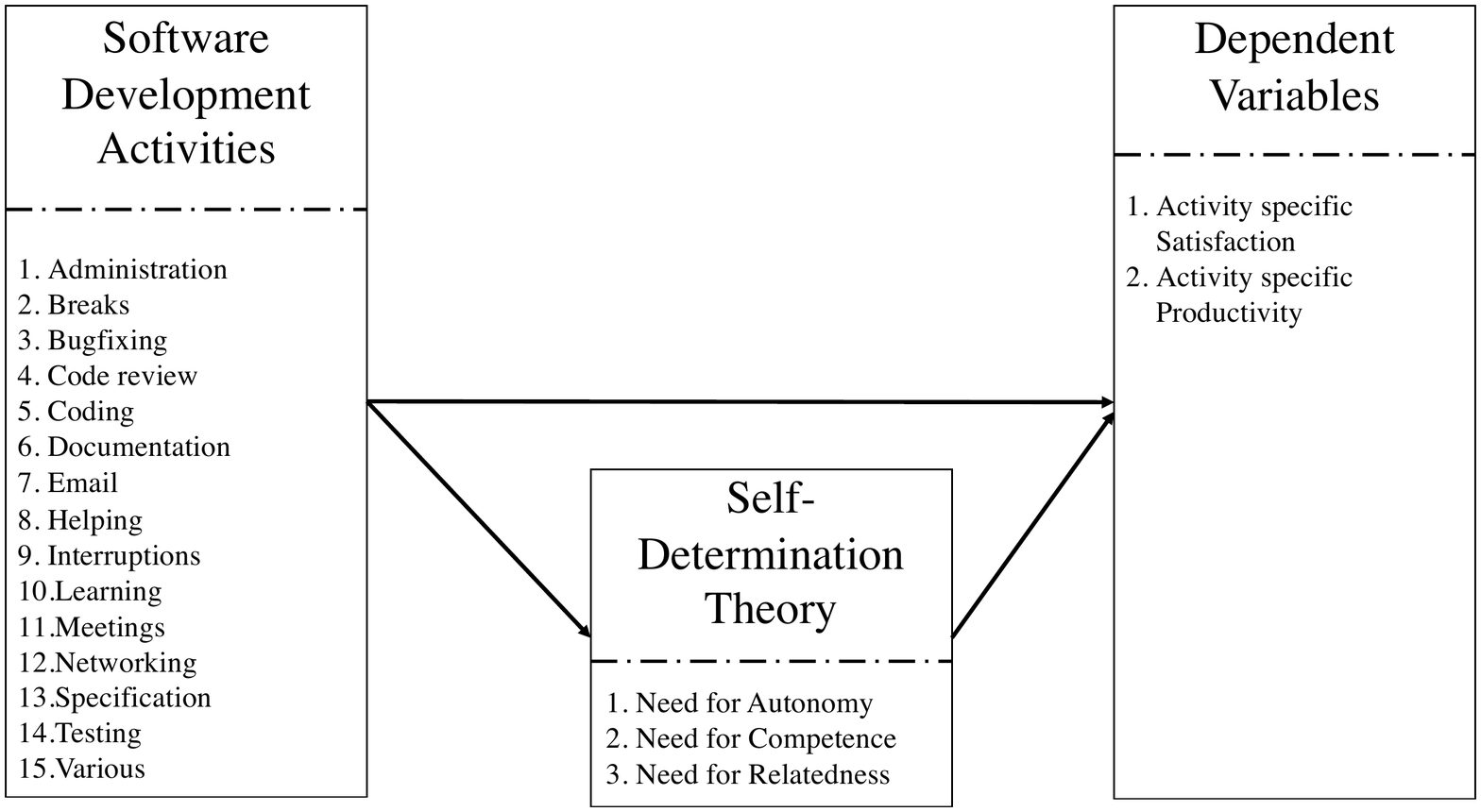}
  \caption{Theoretical Framework of Self-Determination Theory (SDT) in the workplace adapted from Deci et al.~\cite{deci2017self}, where software engineering activities are the workplace-related independent variables, and SDT the mediating variable.}
  \label{fig:theory1}
\end{figure*}

The motivation related to specific job activities influences employees’ productivity and well-being.
Specifically, according to Deci et al., it mediates workplace-specific context such as developers' activity with performance and wellness~\cite{deci2017self}, as depicted in Figure \ref{fig:theory1}.
In other words, the three basic needs of SDT applied to developers' activity should be positively associated with well-being and productivity.

\subsection{Participants}
\label{SSec:Participants}

For the exploratory study, a power analysis using G*Power~\cite{faul2009statistical} version 3.1 revealed that to detect a small-to-medium effect size of $r$ = .20, using a power of $1-\beta$ = .80 (for a two-sided test), a sample size of at least 190 participants is required\footnote{With $r$, we mean Pearson's r, which is a measurement of linear association between two variables; its values ranges between -1 (perfect negative linear association) and +1 (perfect positive linear association). Values around 0 suggest that there is no linear association between two variables. Statistical power is the probability of detecting an effect of at least a given effect size with a certain probability (here: .80).}. We assumed an effect size of $r$ = .20 because this is close to the medium effect size in individual difference research~\cite{gignac2016effect} from which many of our variables stem (e.g., SDT). We used a power of .80 because it is conventional to keep the false-negative rate (i.e., the $\beta$-error) to 1 - .80 = .20 or lower~\cite{cohen1992power}. If we had assumed a larger effect size, fewer participants would have been needed to detect such a larger effect with a power of .80. 

Participants were selected from a broader set of 500 software engineers who were carefully selected through a multistage process in a previous study by Russo \& Stol~\cite{russo2020gender}.
To select this initial pool of participants we applied a three-level screening process. 
First, we \textit{pre-screened} the participants on the Prolific platform.
The initial pre-screening criteria was knowledge of software development techniques, do computer programming for a living, use technology at work, and have an approval rate of 100\% in previous studies.
This left us with 2,897 members candidates.
Then, we performed a \textit{competence screening}.
With the help of a questionnaire, we assessed in a time-boxed fashion the candidates' knowledge with one question about software design and two about programming. 
After this phase, 514 candidates were included in our sample.
Finally, we focused on the candidates' attention with a \textit{quality screening}, where we excluded informants who had a suspicious response pattern or have not passed attention checks of a 10-minutes long questionnaire about personality traits.
The final set contained 483 fully screened software engineers.

For this study, we only selected professionals (from the Russo \& Stol pool) who were working from home during the pandemic and live in countries with comparable lockdown measures.
We used the following criteria: the country had to be in an official lockdown and those measures had to be rather homogeneous across the country.
For example, countries such as Sweden with rather liberal lockdowns were excluded.
Similarly, in Germany individual regions decided whenever the lockdown had to be applied\footnote{The reader should be aware that at the time of the first data collection the understanding among people of COVID-19 and health policies was much different than it is today. For example, we also excluded Denmark who wanted initially to follow Sweden's herd immunity strategy.}.
Finally, we obtained a sample of 192 software engineers who completed the first survey ($M_{age}$~=~36.65 years, $SD$~=~10.77, range~=~19–63; 154 men, 38 women). Of those, 184 participated in the second wave two weeks later. 
We provide demographic information on participants' gender, age, and location in Table~\ref{tab:demographics}.
We collected our data between 26 and 30 April 2020 (wave 1) and between 10 and 13 May 2020 (wave 2). 

To identify participants for the confirmatory study, we also first run a power analysis, which revealed that a sample size of 77 is sufficient to detect a medium effect size with three predictors (i.e., need for autonomy, competence, and relatedness) with a power of .80. However, to keep the length of the survey to a manageable amount, participants only selected three activities they performed during the day. They completed a series of questions that expressly referred to each of the three activities. We therefore aimed to recruit around 300 participants, to obtain for multiple activities the required sample size of at least 77 participants. 
To ensure that the participants were software engineers, we run a pilot study to screen our informants with questions developed by Danilova et al.~\cite{danilova2021you}. 
The survey design is comparable with the previous exploratory one. 
The \textit{pre-screened} followed the same criteria.
What was different is the \textit{competence screening}, where we asked specific questions developed and validated by Danilova et al.
Regarding the \textit{quality screening}, of the 300 selected participants, 10 participants failed at least one test item and/or completed the survey in less than 4 minutes and were excluded. 
The vast majority of participants, 210, worked in `Software \& IT,', 20 in `Education \& Research,' and 11 in `Finance, banking \& insurance.'

To ensure high data quality, we recruited participants from the academic data collection platform Prolific Academic and compensated participants above the US minimum wage~\cite{palan2018prolific,russo2022recruiting}.
The survey was run using Qualtrics.

\begin{table}[]
\caption{Demographic information of both samples. For brevity, we list the most common countries among our sample.}
\centering
\begin{tabular}{@{}lll@{}}
\toprule
                             & \textit{\textbf{N} Exploratory study} & \textit{\textbf{N} Confirmatory study}  \\ \midrule
Sample size                  & 192                         & 290 \\ \midrule
United Kingdom               & 63  (32.8\%)                &  36 (12.4\%) \\
United States                & 52  (27.1\%)                &  22 (7.6\%) \\
Portugal                     & 19  (9.9\%)                 &  54 (18.6\%) \\
Poland                       & 10  (5.2\%)                 &  63 (21.7\%) \\
Italy                        & 7   (3.6\%)                 &  13 (4.5\%) \\
Ireland                      & 5   (2.6\%)                 &  3 (1.0\%) \\ 
Other Europe                 & 23  (12.0\%)                &  61 (21.0\%) \\ 
Other North America          & 8   (4.2\%)                 &  19 (6.6\%) \\ 
Oceania                      & 5   (2.6\%)                 &  6 (2.1\%) \\ 
South America                & 0   (0\%)                   &  4 (1.4\%) \\ 
Africa                       & 0   (0\%)                   &  5 (1.7\%) \\ 
Missing                      & 0   (0\%)                   &  4 (1.4\%) \\ \midrule
Men                          & 154  (80.2\%)               &  242 (83.4\%) \\ \midrule
Women                        & 38  (19.8\%)                &  48 (16.6\%) \\ \midrule
Mean age ($SD$, range)       & 36.7 (10.7, 19-63)          & 25.85 (6.44, 18-60) \\  \bottomrule
\end{tabular}
\label{tab:demographics}
\end{table}

\subsection{Measurements for the exploratory longitudinal study}

For the exploratory study, we derived the variables from a related project. 
For a complete presentation of the used instruments, we directly refer to Russo et al.~\cite{russo2020predictors} and the Supplementary Materials. Most of the scales described below have been cited between hundreds and tens of thousands times and been used across a wide range of contexts (e.g., organizational, clinical). The longitudinal design also allowed us to compute test-retest reliabilities, $r_{it}$ (i.e., the stability of responses across two or more time-points), by correlating responses given by participants at time 1 with those at time 2 (we are using \textit{time} and \textit{wave} interchangeably), which provides additional information about a scale's reliability to the commonly used Cronbach's alpha~\cite{mcdonald2013test}. 
Test-retest reliabilities close to 0 are undesirable since they indicate a low association between the two-time points, suggesting, among others, poor data quality. Cronbach's alpha is a measure of scale reliability. For exploratory research, using new measurement scales, values above .60 are desirable while for confirmatory research the threshold is above .70 (and below .95)~\cite{hair2013multivariate}.

\textbf{Activities.} We measured the same 15 activities that were measured by Meyer et al.~\cite{meyer2019today}. We did this because we believe they covered most activities and to have a pre-pandemic comparison group. We asked participants, "During the past week, how much time did you spend on each task percentage-wise (\%)?" This was followed by the 15 activities, rated on a 101-point slider-scale ranging from 0\% to 100\%. For the activities which might have been more ambiguous, a brief explanation was added in brackets such as `Helping (helping, managing or mentoring people),' `Networking (maintaining relationships).' The 15 activities are coding, bugfixing, meetings, testing, email, breaks, code review, specification, learning, helping, administration, interruptions, documentation, various (i.e., other activities not listed above), and networking.

\textbf{Well-being}. We used the Satisfaction with Life Scale~\cite{diener1985satisfaction}, because it is one of the most validated scales and because it shows good convergent and discriminant validity~\cite{pavot2009review}. Example items validated  include "The conditions of my life in the past week were excellent” and “I was satisfied with my life in the past week". Responses were given on a 7-point scale ranging from 1 (strongly disagree) to 7 (strongly agree). Our Cronbach's alpha values to measure internal consistency for both waves were the following $\alpha_{time 1}~=~.90$, $\alpha_{time 2}~=~.90$ ( $r_{it} = .72, p < .001$). 

\textbf{Productivity}. Measuring productivity in software engineering is a highly debated issue. Some scholars, for example, suggest making the measurement more objective by using function points~\cite{wagner2018systematic}. Ko has criticized this viewpoint as being detrimental in the long run~\cite{ko2019we}.
On the other hand, other researchers propose a self-reflection measure with developers' self-reporting their daily productivity~\cite{meyer2014software}.
In this work, we adopted a similar approach. 
We did not use a standard measure (e.g., such as Ralph et al.~\cite{Ralph2020pandemic} did). 
Instead, we operationalized productivity as a function of time spent working and efficiency per hour, compared to a typical week. Specifically, we asked respondents three items: "How many hours have you been working approximately in the past week?” (Item 1), “How many hours were you expecting to work over the past week assuming there would be no global pandemic and lockdown?” (Item 2), and "If you rate your productivity (i.e., outcome) per hour, has it been more or less over the past week compared to a normal week?” (Item 3). Item 3 measured perceived efficacy and was answered on a bipolar scale that ranged from "100\% less productive" to "100\% more productive", with the scale mid-point being "‘0\%: as productive as normal". We computed productivity with the following formula: productivity $= (Item 1/Item 2) \times ((Item 3 + 100)/100)$. Productivity scores from 0 to .99 would reflect lower than normal productivity, scores of 1 the same amount of productivity, and scores above 1 higher levels of productivity. \par 
The reason for this choice is that we wanted to investigate the variance in productivity while working remotely as compared to being in the office. We acknowledge that some readers might have some concerns with this approach. For example, software engineers might understand productivity differently. While one software engineer might feel productive when having been asked to do a lot of tasks other than their main task for the week with high priority, whereas another software engineer might feel less productive. However, this is an issue of all our scales (e.g., we do not know whether participants interpret/instantiate autonomy, competence, or well-being in the same way), but nevertheless we find strong correlations among these variables. This interpretation is supported from psychological research: There is substantial heterogeneity in how people interpret human values (e.g., equality, freedom, security)~\cite{hanel2018cross}. Nevertheless, values are still strong predictors of personality and beliefs~\cite{kajonius2015hedonism,saroglou2004values}. As long as there is no systematic bias in how our participants understood productivity -- and we do not assume there is -- we do not believe this is an issue. Additionally, test-retest reliability correlation was large, $r_{it} = .50, p < .001$, and productivity correlated negatively with the number of breaks taken (Tab. \ref{tab:activities1}). 

\textbf{Stress}. We used a 4-item version of the Perceived Stress Scale~\cite{cohen1983global}, as it is an often used and well-validated scale~\cite{lee2012review}. Example items include "In the last week, how often have you felt that you were unable to control the important things in your life?” and “In the last week, how often have you felt confident about your ability to handle your personal problems?" The response scale ranged from 1 (Never) to 4 (Very often). $\alpha_1~=~.80$, $\alpha_2~=~.77$ ($r_{it} = .73, p < .001$).

\textbf{Boredom}. We used the Boredom Proneness Scale~\cite{farmer1986boredom,struk2017short}, because it is a well-validated scale~\cite{tam2021boredom}. Example items include "It is easy for me to concentrate on my activities” and “Many things I have to do are repetitive and monotonous". Items were answered on a 4-point scale ranging from 1 (Strongly disagree) to 7 (Strongly agree). $\alpha_1~=~.87$, $\alpha_2~=~.87$, ($r_{it} = .69, p < .001$).

\textbf{Autonomy, competence, and relatedness}. To measure the three needs of the self-determination theory~\cite{ryan2000self}, we used the psychological needs scale~\cite{sheldon2012balanced}, which is also a well-validated scale~\cite{neubauer2016validation}. Example items include “I was free to do things my own way" (need for autonomy), "I did well even at the hard things” (need for competence), and "I felt unappreciated by one or more important people" (need for relatedness, recoded). Need for autonomy's Cronbach's alpha level were: $\alpha_1~=~.72$, $\alpha_2~=~.76$ ($r_{it} = .76, p < .001$); for Competence: $\alpha_1~=~.77$, $\alpha_2~=~.65$ ($r_{it} = .76, p < .001$); and for Relatedness: $\alpha_1~=~.79$, $\alpha_2~=~.78$ ($r_{it} = .71, p < .001$).

\textbf{Quality and quantity of communication with colleagues and line managers}. We used a self-developed three items instrument to capture how positive and supportive the communication has been with colleagues and line managers. The items are “I feel that my colleagues and line manager have been supporting me over the past week”, “I feel that my colleagues and line manager believed in me over the past week”, and “Overall, I am happy with the interactions with my colleagues and line managers over the past week.” ($\alpha_1~=~.88$, $\alpha_2~=~.92$; $r_{it} = .67, p < .001$).

\textbf{Daily Routines}. We developed a five items scale to capture participants' daily habits, as having automaticity in one's life frees cognitive resources for other things such as work~\cite{moors2006automaticity}. The items were designed to capture a broad range of daily activities that were possible during the regulations in most countries at the time of data collection (spring 2020). The items are  “I am planning a daily schedule and follow it”, “I follow certain tasks regularly (such as meditating, going for walks, working in timeslots, etc.)”, “I am getting up and going to bed roughly at the same time every day during the past week”, “I am exercising roughly at the same time (e.g., going for a walk every day at noon)”, and “I am eating roughly at the same time every day” ($\alpha_1~=~.75$, $\alpha_2~=~.78$; $r_{it} = .73, p < .001$).

\textbf{Distractions at home}. We developed a two items scale to measure perceived distraction in general as measuring the exact cause for distractions would have been beyond the scope of our study. The items are “I am often distracted from my work (e.g., noisy neighbors, children who need my attention)” and “I am able to focus on my work for longer time periods” (recoded) ($\alpha_1~=~.64$, $\alpha_2~=~.63$; $r_{it} = .63, p < .001$).

\subsection{Measurements for the confirmatory cross-sectional study}
\subsubsection{Measurement of activity-specific variables}
After providing informed consent, participants were instructed "\textit{Which of the following tasks have you spent most time with yesterday? For example, when you spent most of your time in two meetings, pick the meeting that went longer. Select three tasks}." Participants selected three of the activities we used in Study 1, except breaks, interruptions, and various, which were excluded, leaving 12 activities: Coding (n = 192), bugfixing (111), testing (96), specification (22), reviewing (91), documenting (40), meetings (87), emails (51), helping (33), networking (11), learning (93), and administration (14). Participants then completed 17 items for each task, 8 measuring our two dependent variables, well-being and productivity, and 9 measuring our three independent variables, need for autonomy, competence, and relatedness. 

\textbf{Satisfaction} was measured with a six items we created for the purpose of the study. The items were created to capture positive and negative aspects of satisfaction~\cite{karademas2007positive}. In other words, some items were reversed scored, which might result in lower reliability (e.g., if a participant gives the item only a cursory read) but comes with the advantage of higher validity~\cite{clifton2019managing}. The wording of the six item is "How stressed were you during the task?" (reversed scored), "How many positive emotions have you felt during the task?", "How bored were you during this task?" (reversed scored), "After completing the task, I felt tired" (reversed scored), "Performing this task frustrated me" (reversed scored), and "I felt exhausted after the task" (reversed scored). The reversed scored items were recorded so that higher scores indicated higher well-being. Answers were given on a scale ranging from 1 (Not at all) to 7 (Very). A principal component analysis revealed that the 6 items were loading on one component, with good internal consistency ($\alpha = .80$). 

\textbf{Productivity} was measured with two items we created for the purpose of the study: ``How productive have you been during this task?", which was answered on a scale ranging from 1 (Not at all) to 7 (Very), and ``What percentage of your goals have you reached during $<task>$," which was answered on a 0-100 scale. We created both items as they measure related, yet slightly different aspects of productivity. For example, a software engineer can feel productive but not have reached all of their goals because unexpected issues occurred while working on an activity. If the issues were overcome, the software engineer might feel productive but have not fully reached their goals. Both items were standardized before being averaged ($\alpha = .50$).

To measure the three independent variables, we adapted three items for each of the three needs of the self-determination theory~\cite{ryan2000self} from the balanced measure of psychological needs scale~\cite{sheldon2012balanced}. The scale measures each of the three needs with six items. We selected those items which we judged as best suitable to be adapted for our purpose. We chose three items to get a good balance between brevity and informativeness: For example, if we had measured each need with only two items, we would have ended up with only one if a participant skipped an item as not applicable; conversely, selecting four items per need would have resulted in nine more items (i.e., 3 needs $\times$ 3 activities) for the full survey, thus increasing its length. All items were answered on a 7-point response scale varying from 1 (Not at all) to 7 (Fully) with an 8th option, `Not applicable.' 

\textbf{Need for autonomy} was measured with ``I was really doing what interests me," ``I was free to do things my own way," and ``I had a lot of pressures I could do without when working on the task" (recoded). However, as the last item was uncorrelated with the other two, $r$s = -.00 and -.14, we only combined the first two items ($\alpha$ = .46) into an Autonomy factor and included the last item as a single-item predictor.\footnote{While the three items usually load on the same factor when measured in a non-specific way~\cite{sheldon2012balanced} -- also among software developers see our exploratory study -- in the context of our study people still can feel pressured to do an activity while being able to do things their own way. This apparent paradox is likely familiar to many researchers: They are often free to pick their own research projects but might then feel obliged to complete them because of pressure from their colleagues, from editors, or to advance in their career -- especially if they have chosen to work on too many projects. Also, given that we have adapted the established balanced measure of psychological needs scale~\cite{sheldon2012balanced} and that the internal consistencies for the task-independent variables are good (mostly $.75 \leq \alpha \leq .90$), we believe that the issue at hand is the adaptation that unexpectedly did not work rather than the data quality.}

\textbf{Need for relatedness} was measured with ``I felt close and connected with people working on the same task as me," ``I felt appreciated by one or more people working on the same task as me," and ``I had disagreements or conflicts with people working on the same task as me" (recoded).  However, as the last item was uncorrelated with the other two, $r$s = .09, .06, we only combined the first two items ($\alpha$ = .73) into a relatedness factor and included the last item as a single-item predictor.\footnote{Some participants might have construed `disagreements or conflicts in the context of specific activities as `mild,' which can happen among colleagues one is usually getting along well or even has befriended~\cite{hood2017conflicts}.}

\textbf{Need for competence} was measured with ``I was successfully completing the task," ``I did well even at the hard things," and ``I struggled to complete the task" (recoded; $\alpha$ = .64). 
Thus, instead of the three predictors, we now have five, two of which are single item predictors. While single-item scales are sometimes considered as problematic because of possible low reliability, they are often used in research and -- assuming there is evidence that participants paid attention to the items as evidenced through good internal consistencies of other scales -- can produce meaningful findings~\cite{gebauer2017religiosity,wolf2020measured}. Indeed, the results of the measures with the two single items are in line with expectations (see below).  

\subsubsection{Measurement of task-independent variables}

Additionally, we also included variables that were suggested to be related to our dependent variables from the exploratory investigation.

\textbf{Resilience} was measured with the 6-item Brief Resilience Scale~\cite{smith2008brief}. Participants indicate how much they agreed with statements such as ``I tend to bounce back quickly after hard times" and ``It is hard for me to snap back when something bad happens" (recoded). Responses were given on a $5$-point scale ranging from $1$ (Strongly disagree) to $5$ (Strongly agree; $\alpha~=~.73$). \par
\textbf{Caring leadership} was measured with the 7-item Caring Leadership Scale~\cite{louis2016caring}. Example items include ``My manager develops an atmosphere of caring and trust" and ``I feel free to discuss work problems with my manager without fear of having it used against me later." Responses were given on a $5$-point scale ranging from $1$ (Strongly disagree) to $5$ (Strongly agree; $\alpha~=~.85$).\par
\textbf{Work-life balance} was measured with a 5-item scale. The items from this and the following five scales were provided by Qualtrics and offered to their users~\cite{Qualtrics2021}. After reading the items, we judged them as appropriate measures of the constructs (e.g., work-life balance) they claim to measure. Example items include "My workload is manageable" and ``I have the flexibility I need in my work schedule to meet both work and personal needs." Responses were given on a $5$-point scale ranging from $1$ (Strongly disagree) to $5$ (Strongly agree; $\alpha~=~.84$). \par
\textbf{Empowerment} was measured with a 7-item scale. Example items include ``I am given the opportunity to be involved in decisions that affect me" and ``Employees are encouraged to participate in decisions that affect their work." Responses were given on a $5$-point scale ranging from $1$ (Strongly disagree) to $5$ (Strongly agree; $\alpha~=~.83$). \par
\textbf{Job Enablement} was measured with a 7-item scale. Example items include ``My job is challenging and interesting" and ``My work-from-home workspace allows me to be productive." Responses were given on a $5$-point scale ranging from $1$ (Strongly disagree) to $5$ (Strongly agree; $\alpha~=~.77$). \par
\textbf{Soft company support} was measured with a 3-items, including ``My company is providing me with the necessary software tools to work from home" and ``My company is providing me with the necessary flexibility so that I can work from home properly." Responses were given on a $5$-point scale ranging from $1$ (Strongly disagree) to $5$ (Strongly agree; $\alpha~=~.64$). \par
\textbf{Hard company support} was measured with a 3-items, including ``My company is supportive in providing me the necessary work from home setting (e.g., chair, screen, mouse)." and ``From the start of the lockdown, my company is taking care also of things it didn't do before (e.g., internet bill, electricity bill)." Responses were given on a $5$-point scale ranging from $1$ (Strongly disagree) to $5$ (Strongly agree; $\alpha~=~.76$). \par
\textbf{Recognition} was measured with a 7-item scale. Example items include ``I receive meaningful recognition when I do a good job" and ``My manager values my contribution." Responses were given on a $5$-point scale ranging from $1$ (Strongly disagree) to $5$ (Strongly agree; $\alpha~=~.89$).

\section{Analysis \& Results}
\label{sec:analysis}

In this section we describe which analyses we used to address our five research questions and the results.

\subsection{RQ1: Has the distribution of daily working activities of software engineers changed while WFH during the pandemic as compared to pre-pandemic daily working activities?}

\begin{figure*}
  \centering
  \includegraphics[width=1\textwidth]{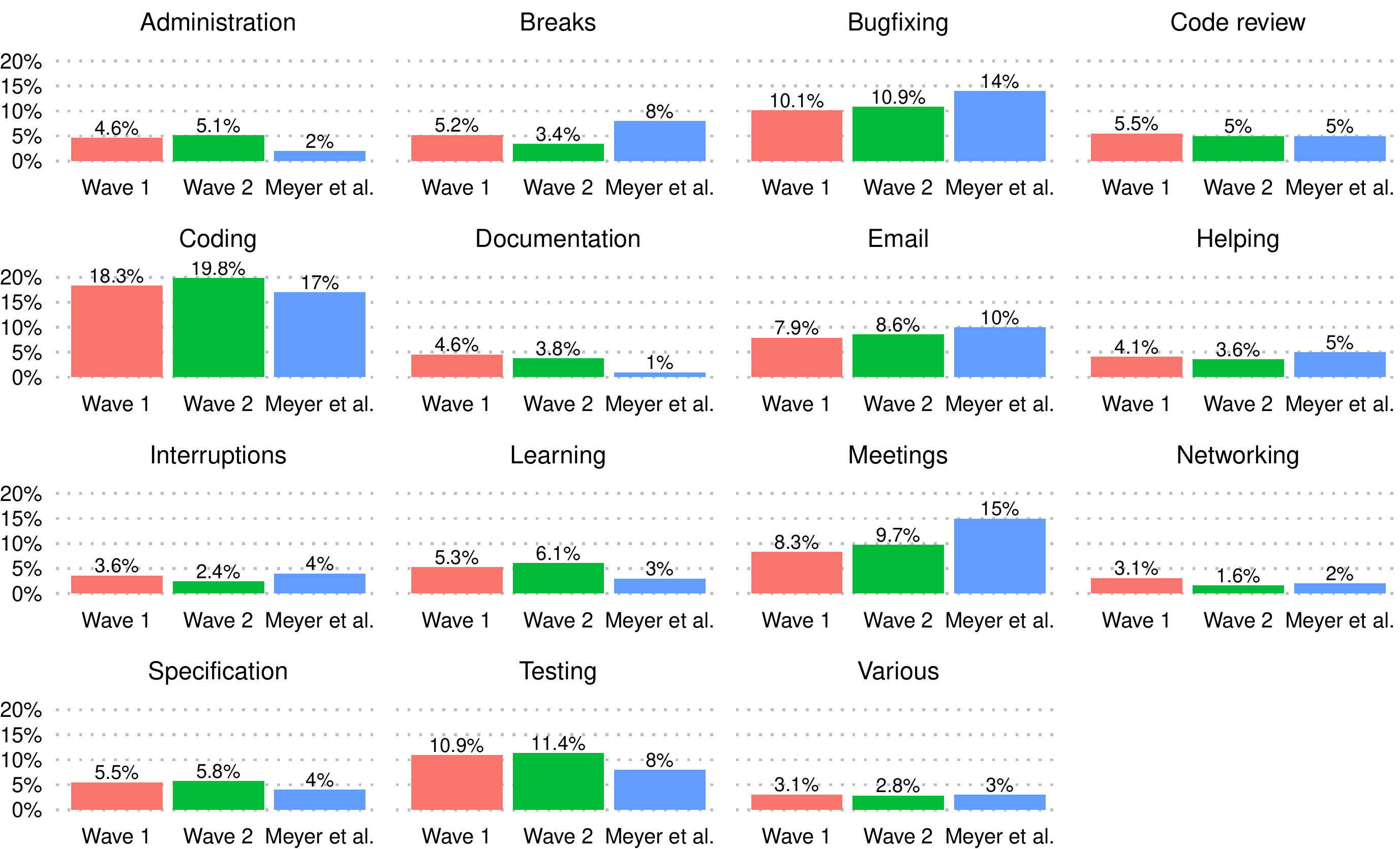}
  \caption{Distribution software engineering work activities during the two waves in our study, and a typical workday of software engineers as reported by Meyer et al.~\cite{meyer2019today}.}
  \label{fig:activities}
\end{figure*}

To test RQ1, we first compared the time participants reported to have spent on each of the 15 activities with those reported by Meyer et al.~\cite{meyer2019today}. The results are displayed in Figure~\ref{fig:activities}, as well as Tables \ref{tab:activities} and \ref{tab:comparison}. To test whether participants in our sample reported spending more or less of their time on certain activities than the software developers surveyed by Meyer et al.~\cite{meyer2019today}, we performed a series of one-sample $t$-tests. For example, we compared the percentages of participants in our sample at time 1 spend coding was significantly different from 15\%, which is the percentage reported by Meyer et al. (see Table~\ref{tab:activities}, second column). That is, we tested whether participants in our sample spend significantly more time (i.e., $>15\%$) or less time ($<15\%$) coding than participants in Meyer et al.'s pre-pandemic study. We performed 15 (activities) $\times$ 2 (time points) = 30 $t$-tests (two-tailed, since we did not have directed hypotheses)\footnote{Because of the large number of comparisons, we adjusted the $\alpha$-threshold from .05 to .003 to reduce the risk of false-positive results. This means that we considered only $p$-values of $< .003$ as statistically significant. This is a standard procedure for studies that involve many variables to ensure reliable results, e.g.,~\cite{hanel2020well}. Note that changing the $\alpha$-threshold impacts the test statistic (e.g., $t-$value), as the test statistic and $p$-value are perfectly associated with any given sample size~\cite{hays_statistics_1994}. For example, for an $\alpha-$threshold of .003 and a sample size of 192 (time 1) or 184 (time 2), the critical $t$-values are 3.006 and 3.008. In other words, only if the $t-$value obtained from a $t-$test is larger than 3.006 (or 3.008), the $p$-value would be $<.003$, and we would consider the test result to be statistically significant. Note that a Bonferroni correction would have resulted in an adjusted alpha-level of .05/30 $\approx$ .0017, which is overly conservative and does not consider that some variables are correlated (e.g., between time 1 and 2). Thus, the adjusted significance threshold of .003 seemed appropriate to us, neither overly conservative nor liberal. Also, we are not interpreting p-values that are just above our threshold. Doing this would be equal to stating that there is a trend towards significance, implying that with a larger sample the effect would have become statistically significant. However, this is not the case~\cite{wood2014trap}}. 
 \par

Software engineers in our sample reported on average to have spent less time bugfixing, in meetings, getting interrupted (only at time 2), helping (only at time 2), and taking breaks; but more time on testing, specification, writing documentation, networking (only at time 1), learning, and administrative activities compared to the participants surveyed by Meyer et al. (Table~\ref{tab:activities}). However, the differences between what our participants and those of Meyer et al. reported differed by only a few percent (see Figure~\ref{fig:activities}). This visual inspection of the data is supported by correlation analysis. The percentages of time spent on the 15 activities\footnote{For the correlations, the Degrees of Freedom are $N-2=13$ with N = 15 activities.} reported by Meyer et al. correlated with $r(13) = .84, p < .0001$ at time 1 and with $r(13) = .83, p = .0001$ at time 2. To obtain those correlations, we correlated the mean percentages reported in columns 2-4 of Table~\ref{tab:activities} with each other. That is, we tested whether the average percentages spent on each activity reported by the participants in the Meyer et al. sample would align with those reported by the participants in our sample at waves 1 and 2. This suggests that while there are some deviations, the overall order of activities remains stable. It further supports the quality of our data. If our participants had responded carelessly or even randomly, those two correlation coefficients would be around 0.

In the next step, we explored whether participants' activities changed over time during the lockdown. To do this, we performed a series of paired $t$-tests (Table~\ref{tab:comparison}). The only statistically significant differences were observed for networking and taking breaks. At time 2, participants spent less time networking and taking breaks compared to time 1. Overall, the relative order of the activities remained very stable across time on the group level (i.e., when correlating the group averages for the activities of time 1 and 2), $r(13) = .99, p < .0001.$

\begin{table}[]
\centering
\sisetup{
    group-digits=true,
    detect-weight=true,
    detect-shape=true,
    table-format=2.2,
    table-alignment=left
}
\caption{Comparisons of both waves with time spend on activities as reported by Meyer et al.~\cite{meyer2019today}}
\resizebox{\textwidth}{!}{%
\begin{tabular}{@{}lrrrS[table-format = -1.3]S[table-format = -1.3]S[table-format = <1.3]S[table-format = <1.3]@{}}
\toprule
\textbf{Activity}       & \textbf{Meyer et al.} & \textbf{M\textsubscript{t1}}    & \textbf{M\textsubscript{t2}}    & \textbf{$t$-value 1} & \textbf{$t$-value 2} & \textbf{$p$1}               & \textbf{$p$2}               \\ \midrule
Coding          & 17\%     & 18.11\% & 19.85\% & 0.901       & 1.89        & 0.369          & 0.060            \\
 Bugfixing      & 14\%     & 10.27\% & 10.85\% & -5.309      & -3.546      & <0.001 $\downarrow$ & <0.001 $\downarrow$ \\
 Meetings       & 15\%     & 8.45\%  & 9.74\%  & -9.951      & -6.628      & <0.001 $\downarrow$ & <0.001 $\downarrow$      \\
 Testing        & 8\%      & 10.96\% & 11.36\% & 3.413       & 3.321       & < 0.001 $\uparrow$  & 0.001  $\uparrow$ \\
 Email          & 10\%     & 7.93\%  & 8.59\%  & -3.686      & -1.584      & <0.001             & 0.115            \\
 Breaks         & 8\%      & 5.21\%  & 3.40\%  & -7.391      & -14.297     & <0.001 $\downarrow$ & <0.001 $\downarrow$ \\
Code review     & 5\%      & 5.44\%  & 5.01\%  & 0.878       & 0.019       & 0.381              & 0.985            \\
 Specification  & 3\%      & 5.49\%  & 5.76\%  & 4.653       & 4.048       & <0.001 $\uparrow$  & <0.001 $\uparrow$ \\
 Learning       & 3\%      & 5.30\%  & 6.07\%  & 4.242       & 3.377       & <0.001 $\uparrow$  & 0.001 $\uparrow$   \\
Helping         & 5\%      & 4.25\%  & 3.60\%  & -2.126      & -3.064      & 0.035              & 0.003  $\downarrow$ \\
 Administration & 2\%      & 4.70\%  & 5.15\%  & 4.575       & 4.279       & <0.001 $\uparrow$  & <0.001 $\uparrow$ \\
Interruptions   & 4\%      & 3.58\%  & 2.42\%  & -1.188      & -5.388      & 0.236              & <0.001 $\downarrow$       \\
 Documentation  & 1\%      & 4.69\%  & 3.77\%  & 5.178       & 5.073       & <0.001 $\uparrow$  & <0.001 $\uparrow$ \\
Various         & 3\%      & 3.17\%  & 2.84\%  & 0.592       & -0.346      & 0.554              & 0.729            \\
Networking      & 2\%      & 3.10\%  & 1.60\%  & 3.040        & -1.485      & 0.003 $\uparrow$  & 0.139            \\ \bottomrule
\end{tabular}
}
\\ \vspace*{3mm}
\footnotesize \textit{Note}. Activity percentages as per `typical workday' following Meyer et al.~\cite{meyer2019today}. M\textsubscript{t1}: mean at time 1 (see also Table~\ref{tab:comparison}), $t$-value 1: $t$-value of one-sample $t$-test from time 1 vs value reported by Meyer et al., p1: $p$-value of one-sample $t$-test from time 1. $\uparrow$ and $\downarrow$ indicate a significant increase or decrease in time spent on activity as compared to~\cite{meyer2019today}. \\
\label{tab:activities}
\end{table}

\begin{table}[]
\centering
\sisetup{
    group-digits=true,
    detect-weight=true,
    detect-shape=true,
    table-format=-1.3,
    table-alignment = left
}
\caption{Comparisons of activities between time 1 and time 2}
\resizebox{\textwidth}{!}{%
\begin{tabular}{@{}lrrrrSS[table-format = <1.3]SS[table-format = 2]S[table-format = 2]S[table-format = 2]@{}}
\toprule
 &  \multicolumn{2}{l}{\textbf{Time 1}} &  \multicolumn{2}{l}{\textbf{Time 2}} &   &   &   &   &   &   \\
 & \textbf{M} & \textbf{SD} & \textbf{M} & \textbf{SD} & \textbf{$t$} & \textbf{$p$} & \textbf{Cohen’s d} & \textbf{Increase} & \textbf{Decrease} & \textbf{Equal} \\ \midrule
Coding          &  18.11\% &  16.973\%    &  19.85\%  &  20.444\% &  -1.502 &  0.135 &  -0.108 &  94 &  74 &  15 \\
Bugfixing       &  10.27\% &  9.722\%     &  10.85\%  &  12.038\% &  -0.422 &  0.673 &  -0.037 &  68 &  86 &  29 \\
Meetings        &  8.45\%  &  9.103\%     &  9.74\%   &  10.767\% &  -2.418 &  0.017 &  -0.153 &  78 &  69 &  36 \\
Testing         &  10.96\% &  11.970\%    &  11.36\%  &  13.720\% &  -0.205 &  0.838 &  -0.014 &  74 &  85 &  24 \\
Email           &  7.93\%  &  7.776\%     &  8.59\%   &  12.103\% &  -0.705 &  0.482 &  -0.063 &  72 &  85 &  27 \\
 Breaks         &  5.21\%  &  5.208\%     &  3.40\%   &  4.362\% &  4.705 &  <0.001 $\downarrow$ &  0.367 &  47 &  102 &  33 \\
Code review     &  5.44\%  &  6.967\%     &  5.01\%   &  7.924\% &  0.385 &  0.700 &  0.035 &  56 &  76 &  50 \\
Specification   &  5.49\%  &  7.407\%     &  5.76\%   &  9.251\% &  -0.194 &  0.847 &  -0.016 &  54 &  68 &  61 \\
Learning        &  5.30\%  &  7.459\%     &  6.07\%   &  12.313\% &  -1.046 &  0.297 &  -0.089 &  51 &  76 &  55 \\
Helping         &  4.25\%  &  4.872\%     &  3.60\%   &  6.184\% &  1.664 &  0.098 &  0.128 &  46 &  81 &  57 \\
Administration  &  4.70\%  &  8.143\%     &  5.15\%   &  9.976\% &  -0.706 &  0.481 &  -0.051 &  55 &  80 &  47 \\
Interruptions   &  3.58\%  &  4.811\%     &  2.42\%   &  3.981\% &  2.814 &  0.005 &  0.263 &  39 &  79 &  62 \\
Documentation   &  4.69\%  &  9.841\%     &  3.77\%   &  7.411\% &  1.256 &  0.211 &  0.116 &  50 &  71 &  62 \\
Various         &  3.17\%  &  3.974\%     &  2.84\%   &  6.384\% &  0.590 &  0.556 &  0.051 &  49 &  78 &  56 \\ 
Networking      &  3.10\%  &  4.977\%     &  1.60\%   &  3.674\% &  4.334 &  <0.001 $\downarrow$ &  0.350 &  31 &  77 &  74 \\ \bottomrule
\end{tabular}
}
\\ \vspace*{3mm}
\footnotesize \textit{Note}. t: $t$-value of a dependent sample $t$-test; Cohen's d: standardized mean difference; Increase: Participants who spend more time on an activity at time 2 compared to time 1; Decrease: Participants who spend less time on an activity; Equal: Number of participants whose score has not changed. $\downarrow$ indicates a significant decrease between time 1 and 2.
\label{tab:comparison}
\end{table}

\subsection{RQ2: Is the distribution of daily working activities related to well-being, productivity, and other variables?}
To test RQ2, we correlated the time participants spent on each activity with the selected variables. This was possible because the activities were mostly uncorrelated in both time points on an individual level. We report Pearson's correlation coefficients ($r$) in our tables since most of the data were normally distributed. However, for the sake of completeness, we also ran a non-parametric Spearman's rank correlations test (reported in the Supplementary Material), which provided us with very similar results, suggesting the robustness of our results.
In total, we computed at both time points 13 (well-being related variables and productivity) $\times$ 15 (activities) = 195 correlations. Given a large number of comparisons, we changed our significance threshold from $\alpha = .05$ to .0005. Again, a Bonferroni correction would have resulted in an adjusted alpha level of .00017, which is overly conservative and does not consider that some variables are correlated (e.g., distractions and stress). Thus, the adjusted significance threshold of .0005 seemed appropriate to us, neither overly conservative nor liberal. This new threshold implies that only correlation coefficients of $r \geq .25$ are significant. This is because the $p$-value of $r = .25$ is just below the .0005 threshold for our sample size of 192, $p \approx .00047$.  \par

\begin{table}[]
\centering
\sisetup{
    group-digits=true,
    detect-weight=true,
    detect-shape=true,
    table-format=-1.2,
    table-alignment = left
}
\caption{Correlations between activities and variables at Time 1}
\resizebox{\textwidth}{!}{%
\begin{tabular}{@{}lSSSSSSSSSS@{}}
\toprule
 &
  {Well being} &
  {Productivity} &
  {Stress} &
  {Boredom} &
  {Relatedness} &
  {Competence} &
  {Autonomy} &
  {Communication} &
  {Daily routines} &
  {Distractions} \\ \midrule
Coding         & 0.09  & -0.02 & -0.20  & -0.04 & 0.15  & 0.13  & 0.18  & 0.08  & 0.11  & -0.10  \\
Bugfixing      & 0.03  & 0.09  & -0.11 & -0.14 & -0.04 & 0.03  & 0.02  & 0.08  & 0.03  & 0     \\
Meetings       & -0.08 & 0.13  & 0.14  & 0.01  & -0.11 & -0.02 & -0.25 & 0     & -0.07 & -0.05 \\
Testing        & -0.02 & -0.01 & -0.04 & -0.06 & 0.13  & 0.06  & -0.02 & -0.06 & 0.15  & -0.02 \\
Email          & -0.08 & 0.12  & 0.04  & -0.05 & -0.07 & -0.05 & -0.05 & 0.06  & 0     & -0.02 \\
Breaks         & 0     & -0.30  & 0.14  & 0.17  & -0.07 & -0.18 & 0.01  & -0.10  & -0.07 & 0.13  \\
Code review    & 0.13  & 0.08  & -0.11 & -0.03 & 0.04  & 0.17  & 0.06  & 0.12  & 0.11  & -0.11 \\
Specification  & 0     & 0.09  & 0.05  & 0.02  & -0.03 & -0.11 & -0.12 & -0.01 & -0.05 & 0.11  \\
Learning       & -0.07 & -0.07 & 0.13  & 0.12  & -0.05 & -0.11 & 0.05  & 0.06  & -0.15 & 0.11  \\
Helping        & 0.07  & 0.10   & -0.08 & -0.12 & 0     & 0.12  & -0.02 & 0.03  & 0     & -0.14 \\
Administration & 0.03  & -0.11 & -0.02 & -0.01 & 0.02  & 0.01  & 0.06  & -0.14 & -0.05 & 0.07  \\
Interruptions  & -0.21 & 0     & 0.20   & 0.07  & -0.27 & -0.21 & -0.20  & -0.08 & -0.21 & 0.12  \\
Documentation  & -0.03 & -0.07 & 0.09  & 0.05  & -0.03 & -0.05 & -0.01 & -0.07 & -0.01 & 0.03  \\
Various        & -0.08 & -0.11 & 0.07  & 0.02  & -0.03 & -0.08 & -0.04 & -0.06 & -0.11 & 0.13  \\
Networking     & 0.06  & 0.08  & 0.10   & 0.15  & 0.03  & 0.07  & 0.07  & -0.02 & -0.03 & -0.05 \\ \bottomrule
\end{tabular}
}
\label{tab:activities1}
\end{table}

\begin{table}[]
\centering
\sisetup{
    group-digits=true,
    detect-weight=true,
    detect-shape=true,
    table-format=-1.2,
    table-alignment = left
}
\caption{Correlations between activities and variables at Time 2}
\resizebox{\textwidth}{!}{%
\begin{tabular}{@{}lSSSSSSSSSS@{}}
\toprule
 &
  {Well being} &
  {Productivity} &
  {Stress} &
  {Boredom} &
  {Relatedness} &
  {Competence} &
  {Autonomy} &
  {Communication} &
  {Daily routines} &
  {Distractions} \\ \midrule
Coding         & 0.11  & 0.02  & -0.07 & 0.01  & 0.14  & 0.08  & 0.19  & 0.12  & 0.13  & 0     \\
Bugfixing      & 0.07  & 0.15  & -0.07 & -0.02 & -0.01 & 0.01  & 0.06  & 0.05  & 0.12  & -0.03 \\
Meetings       & -0.09 & 0     & 0.02  & -0.02 & -0.03 & -0.01 & -0.17 & 0.01  & -0.03 & -0.02 \\
Testing        & 0.03  & 0.07  & 0.04  & -0.08 & 0.08  & -0.02 & -0.02 & 0.01  & 0     & -0.09 \\
Email          & -0.13 & -0.06 & 0.01  & 0.03  & -0.09 & -0.05 & 0.01  & -0.21 & -0.10  & 0.05  \\
Breaks         & -0.11 & -0.16 & 0.03  & 0.16  & -0.09 & -0.15 & -0.01 & -0.08 & -0.02 & 0.07  \\
Code review    & -0.02 & -0.05 & 0.07  & 0.11  & -0.01 & -0.05 & -0.14 & -0.09 & -0.07 & 0.03  \\
Specification  & 0     & 0.09  & 0.03  & 0.10   & -0.12 & -0.01 & -0.10  & 0.18  & -0.02 & 0.01  \\
Learning       & 0.03  & -0.21 & 0.06  & 0.03  & 0.06  & -0.01 & 0.06  & -0.01 & -0.06 & 0.17  \\
Helping        & 0.01  & 0.03  & -0.11 & -0.19 & 0.16  & 0.13  & 0.02  & 0.12  & 0.01  & -0.13 \\
Administration & -0.09 & -0.05 & 0.09  & -0.02 & -0.10  & -0.04 & -0.11 & -0.18 & -0.10  & 0.03  \\
Interruptions  & -0.08 & 0.04  & 0.05  & -0.05 & -0.04 & -0.02 & -0.06 & 0     & 0.03  & -0.05 \\
Documentation  & 0.01  & 0.13  & -0.03 & -0.04 & -0.13 & 0.02  & -0.05 & -0.05 & -0.07 & -0.01 \\
Various        & 0.03  & -0.03 & -0.01 & 0.09  & -0.11 & -0.06 & 0.03  & -0.03 & 0.02  & -0.05 \\
Networking     & 0.04  & -0.07 & -0.13 & -0.13 & 0.07  & 0.08  & 0.05  & 0.10   & 0.06  & -0.11 \\ \bottomrule
\end{tabular}
}
\label{tab:activities2}
\end{table}

The correlation coefficients are presented in Table~\ref{tab:activities1} and Table~\ref{tab:activities2}. 
This analysis did not show substantially significant results across both time points at $\alpha = $.0005. 
At time 1, three significant correlations emerged which were at time 2 no longer significant. First, productivity was negatively correlated with time spent on breaks, $r = -.30, p = .00002$, which can be considered as a further validation of our productivity measure rather than a meaningful finding itself. However, the correlation between productivity and time spent on breaks was again negative but did not reach statistical significance, $r = -.16, p = .03$.
Second, relatedness correlated negatively with interruptions at time 1, $r = -.27, p = .0002$, but not at time 2, $r = -.04, p = .58$. Third, autonomy correlated negatively with meetings at time 1, $r = -.25, p = .00048$, but not at time 2, $r = -.17, p = .02.$
Overall, we conclude that work activities carried out at home are not related to well-being, productivity, and other variables.


\subsection{RQ3: Do the needs for autonomy, competence, and relatedness predict software engineers' activity-specific satisfaction and productivity?}
\begin{figure}[h!]
    \centering
    \includegraphics[width=.9\textwidth]{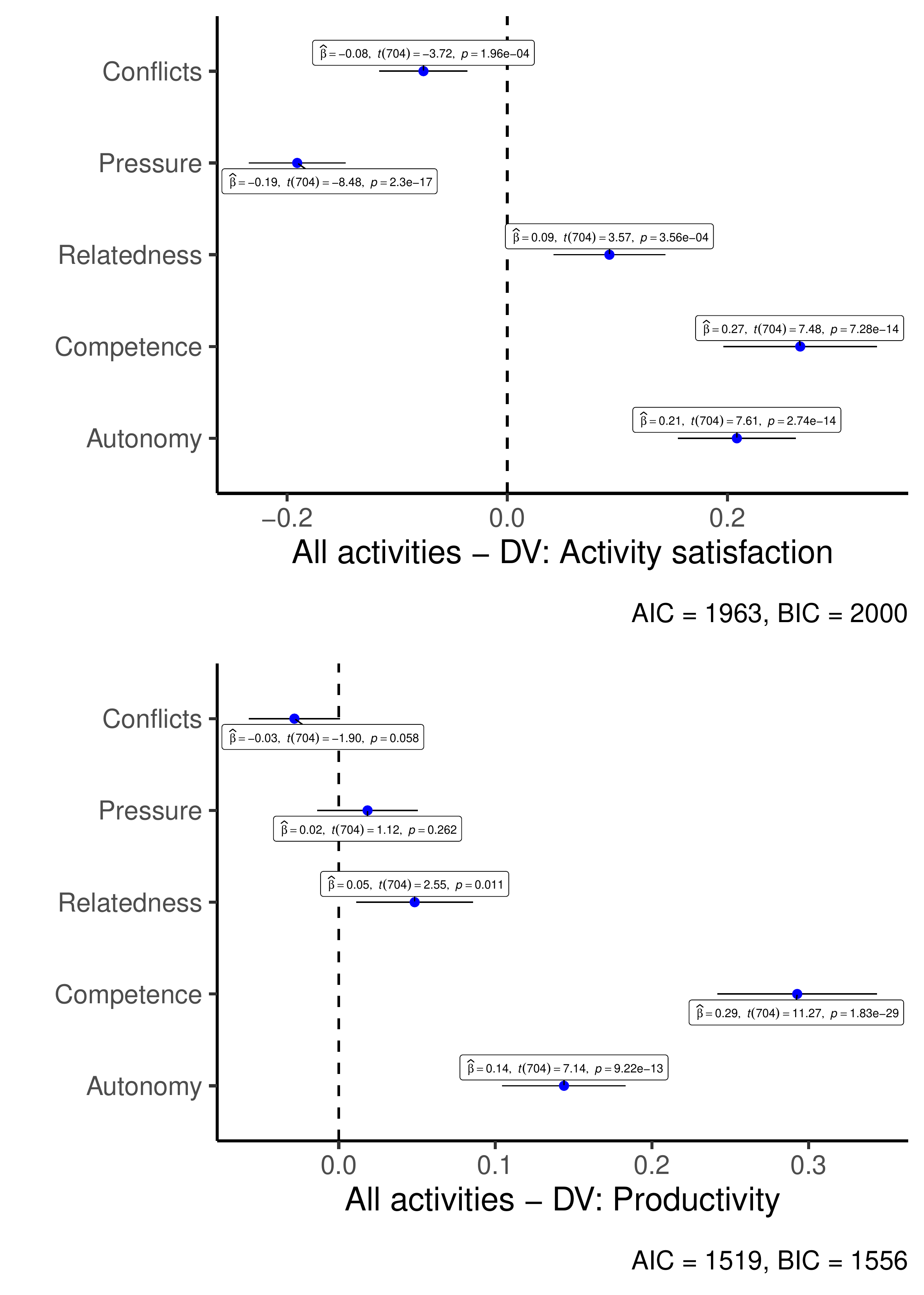}
    \caption{Predictors of activity satisfaction and productivity across all activities. The horizontal lines represent 95\%-CIs.}
    \label{fig:coeff}
\end{figure}

To test the third research question, we run in a first step two linear-mixed models with random intercepts across all activities using the R-package lme4, version 1.1-25~\cite{Bates2015fitting}. A linear-mixed model is superior to a standard multiple linear regression because the responses are not independent, which is an assumption of regression analysis~\cite{brauer2018linear}. Each participant responded to three activities, making them dependent. Ignoring dependencies can result in biases such as an inflated type-I error rate (i.e., false positives)~\cite{judd2012treating}. 
Figure~\ref{fig:coeff} displays the results. Across all activities, activity satisfaction was negatively predicted by conflicts and pressure, and positively by autonomy, competence, and relatedness\footnote{All graphs were created using the R-packages ggplot2, version 3.3.2~\cite{Wickham2016ggplot2}, and ggstatsplot, version 0.6.1~\cite{patil2021visualizations}.}. In contrast, productivity was only predicted by autonomy, relatedness, and especially competence. 

\begin{figure}[]
    \centering
    \includegraphics[width=1\textwidth]{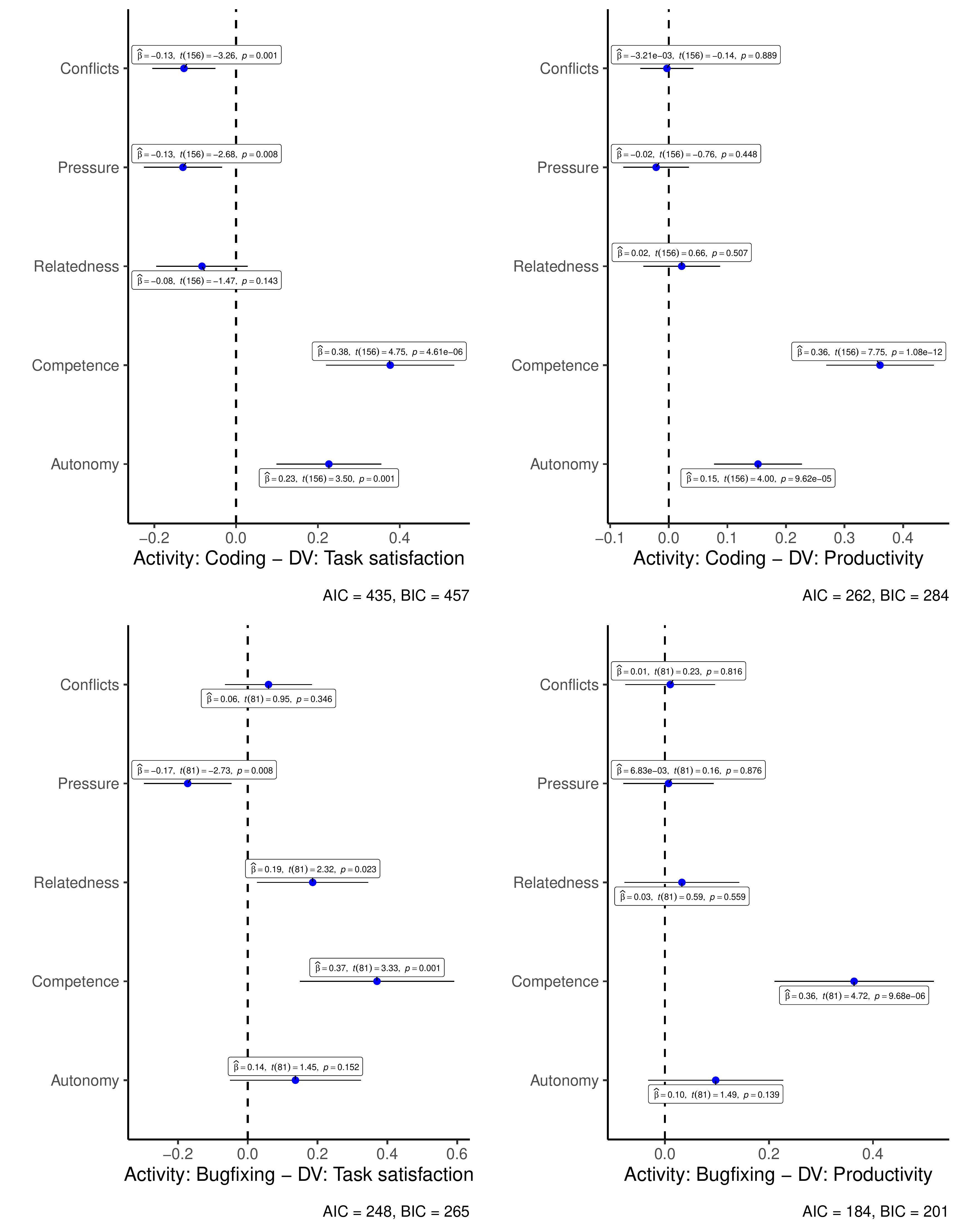}
    \caption{Predictors of well-being and productivity across activities with $n \geq 77$. The horizontal lines represent 95\%-CIs. Plot one of three.}
    \label{fig:coeff2}
\end{figure}

In the next step, we tested whether the pattern of our findings would hold within each of the completed activities by at least 77 participants. This threshold was used because the power analysis reported above revealed that at least 77 participants were needed to detect a medium effect size. As can be seen in Figures~\ref{fig:coeff2} to~\ref{fig:coeff2-2}, the pattern of the result was mostly consistent across the activities, but some minor deviations occurred. For example, for meetings, competence did not matter for participant's activity satisfaction and productivity, but autonomy mattered. In other words, during meetings, it matters more whether people have the feeling they are autonomous rather than competent.


\begin{figure}[]
    \centering
    \includegraphics[width=1\textwidth]{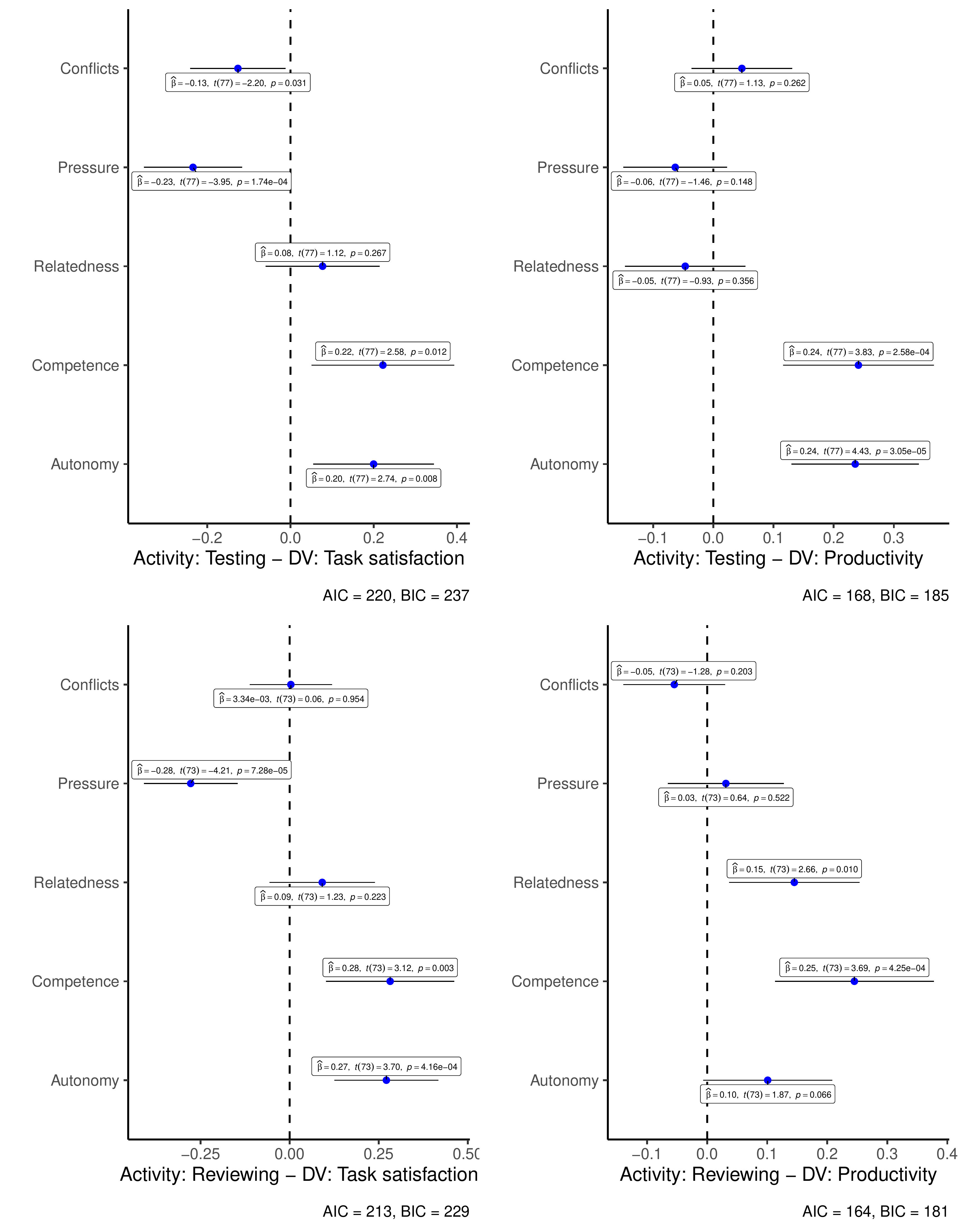}    
    \caption{Predictors of well-being and productivity across activities with $n \geq 77$. The horizontal lines represent 95\%-CIs. Plot one of three.}
    \label{fig:coeff2-2}
\end{figure}

\begin{figure}[]
    \centering
    \includegraphics[width=1\textwidth]{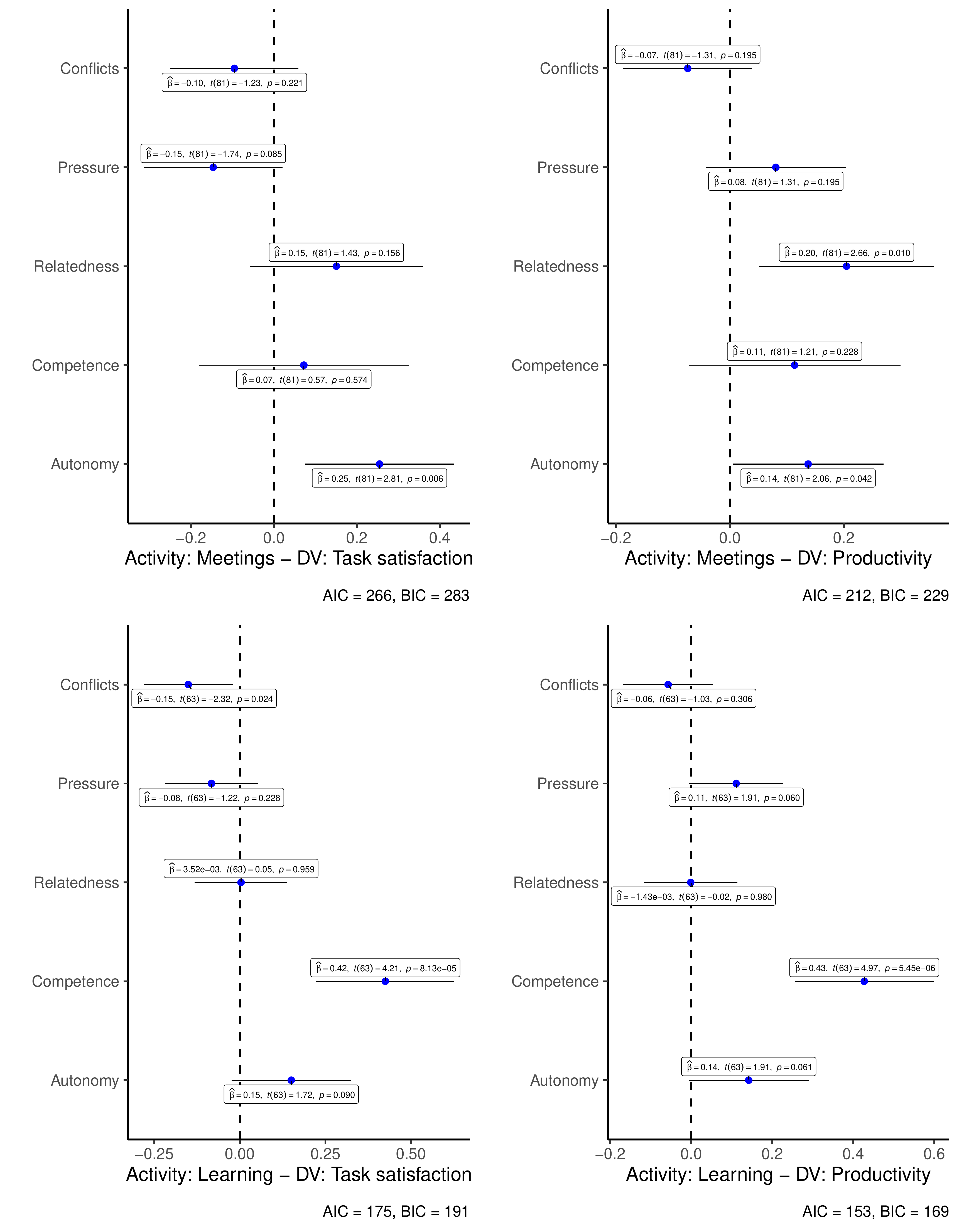}    
    \caption{Predictors of well-being and productivity across activities with $n \geq 77$. The horizontal lines represent 95\%-CIs. Plot one of three.}
    \label{fig:coeff2-2}
\end{figure}

\subsection{RQ4: Are the associations between activity satisfaction and productivity moderated by resilience and company support? }
We tested the fourth research question by running a series of 2 (DV: activity satisfaction vs. productivity) $\times$ 5 (IVs: activity-specific variables autonomy, competence, relatedness, conflict, pressure) $\times$ 8 (moderators: resilience, leadership, balance, empowerment, enablement, soft-support, hard-support, recognition) = 80 moderated regression analyses. Specifically, we multiplied each of the task-dependent variables with each of the task-independent variables. Given a large number of tests, we set our $\alpha$-level to .001 to reduce the likelihood of false-positive results. However, none of the interactions reached statistical significance, $ps > .001$. Together, this suggests that only activity-specific variables matter for activity satisfaction and productivity. \par
Additionally, we tested whether any of the seven task-independent variables would be associated with activity satisfaction and productivity; we again run two linear-mixed models with random intercepts across all activities. The predictors were resilience, leadership, balance, empowerment, enablement, soft support, hard support, and recognition. None of the predictors reached statistical significance, $p > .16$. \par

\subsection{RQ5: Do software engineers' work activities while WFH during the pandemic affect their activity-specific well-being, productivity, and psychological needs?}
\begin{figure}
    \centering
    \includegraphics[width=.85\textwidth]{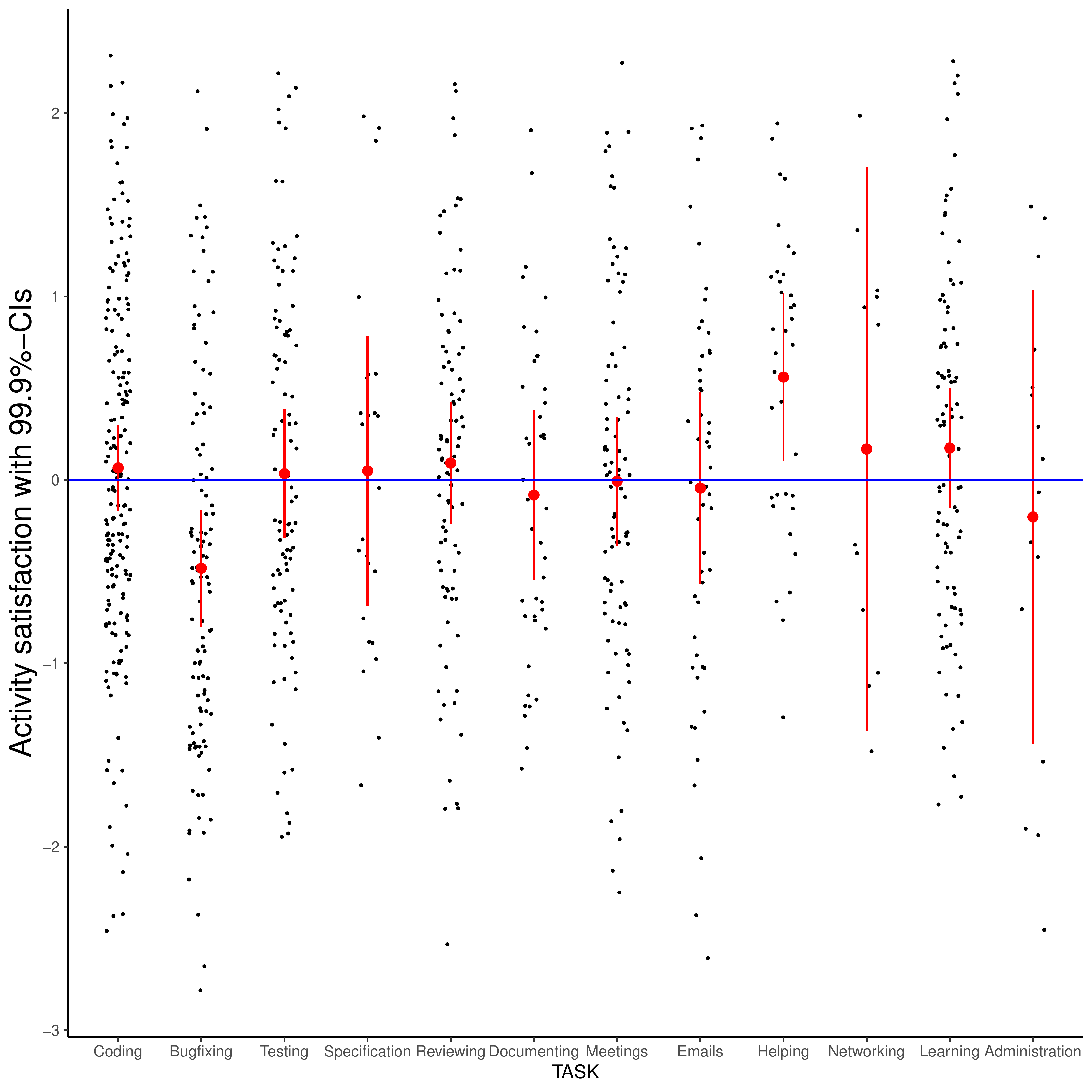}
    \caption{Differences between activities regarding activity satisfaction. Red lines represent 99.9\%-CIs.}
    \label{fig:tasks1}
\end{figure}

Since our design had left many empty cells\footnote{Please recall that participants only responded to the top 3 activities out of a total of 12 possible options, as per survey design.}, a standard approach such as a within-subject ANOVA was not possible (e.g., no participant reported that they were networking and doing administrative activities). We therefore standardized all of our seven outcome variables and tested whether activities would lie above or below the mean for each scale using a series of one-sample t-tests. This approach allows testing whether doing a specific activity increases or decreases, for example, activity satisfaction compared to the average of all activities. Considering the high number involved in our analysis, we set the new alpha-level to .001, which means that we will only consider results to be significant if $p <$ .001 or the 99.9\%-CI does not include zero. 
Results are displayed in Figures~\ref{fig:tasks1} and \ref{fig:tasks2} and Tables~\ref{tab:a} and~\ref{tab:b}. Activity satisfaction was on average lower when participants were bugfixing [$M$ = -0.48, $SD$ = 1.02, $t(114)$ = -5.07, $p < .0001$], and higher when participants were helping others [$M$ = 0.56, $SD$ = 0.77, $t(35)$ = 4.39, $p = .0001$]. Further, participants experienced higher levels of autonomy when coding and lower levels of autonomy when being in meetings and writing emails. Competence was lower when bugfixing and higher when helping people. Relatedness was only higher when people were helping. Pressure and conflict were not impacted by task.

\begin{table}[]
\caption{Differences between activities}
\label{tab:a}
\resizebox{\textwidth}{!}{%
\begin{threeparttable}[t]
\sisetup{
    group-digits=true,
    detect-weight=true,
    detect-shape=true,
    table-format=-1.4,
    table-alignment = left
}
\begin{tabular}{@{}lSSSSSSSSSSSSSSSS@{}}
\toprule
               & \multicolumn{4}{l}{\textbf{Activity satisfaction}}   & \multicolumn{4}{l}{\textbf{Activity productivity}}   & \multicolumn{4}{l}{\textbf{Autonomy}}              & \multicolumn{4}{l}{\textbf{Competence}}            \\
               & \textit{M} & \textit{SD} & \textit{t} & \textit{p} & \textit{M} & \textit{SD} & \textit{t} & \textit{p} & \textit{M} & \textit{SD} & \textit{t} & \textit{p} & \textit{M} & \textit{SD} & \textit{t} & \textit{p} \\ \midrule
Coding         & 0.066      & 0.973       & 0.945      & .346       & 0.084      & 0.895       & 1.304      & .1938      & 0.299      & 0.901       & 4.607      & 0          & -0.133     & 0.921       & -2.015     & .0453      \\
Bugfixing      & -0.481     & 1.016       & -5.073     & 0          & -0.056     & 1.009       & -0.59      & .5562      & -0.093     & 0.86        & -1.151     & .2522      & -0.355     & 1           & -3.804     & .0002      \\
Testing        & 0.035      & 1.016       & 0.334      & .7388      & 0.157      & 0.94        & 1.65       & .1023      & 0.023      & 1.09        & 0.205      & .8378      & 0.031      & 1.031       & 0.298      & .7666      \\
Specification  & 0.05       & 0.981       & 0.253      & .8021      & 0.347      & 0.758       & 2.289      & .0312      & 0.263      & 0.815       & 1.614      & .1196      & 0.011      & 1.191       & 0.045      & .9642      \\
Reviewing      & 0.092      & 0.935       & 0.953      & .3429      & 0.235      & 0.947       & 2.391      & .0189      & -0.024     & 1.013       & -0.225     & .8226      & 0.152      & 1.074       & 1.363      & .1763      \\
Documenting    & -0.082     & 0.86        & -0.625     & .5355      & 0.06       & 0.878       & 0.447      & .6574      & -0.294     & 1.018       & -1.869     & .0687      & 0.208      & 0.78        & 1.752      & .0871      \\
Meetings       & -0.006     & 0.982       & -0.062     & .9505      & -0.284     & 1.104       & -2.456     & .016       & -0.397     & 1.032       & -3.674     & .0004      & 0.127      & 0.926       & 1.303      & .1958      \\
Emails         & -0.044     & 1.084       & -0.295     & .7688      & -0.017     & 1.051       & -0.113     & .9101      & -0.553     & 1.052       & -3.788     & .0004      & 0.382      & 0.937       & 2.943      & .0049      \\
Helping        & 0.561      & 0.766       & 4.391      & .0001      & 0.202      & 1.196       & 1.015      & .3173      & 0.237      & 0.766       & 1.857      & .0718      & 0.571      & 0.727       & 4.714      & 0          \\
Networking     & 0.169      & 1.199       & 0.487      & .6356      & -0.946     & 1.004       & -3.264     & .0075      & -0.06      & 1.199       & -0.173     & .866       & 0.026      & 0.699       & 0.13       & .8991      \\
Learning       & 0.174      & 0.948       & 1.802      & .0747      & -0.22      & 1.031       & -2.086     & .0396      & 0.236      & 0.885       & 2.603      & .0108      & -0.071     & 1.085       & -0.645     & .5203      \\
Administration & -0.202     & 1.216       & -0.663     & .5172      & -0.395     & 1.09        & -1.451     & .1674      & -0.412     & 1.468       & -1.122     & .2796      & -0.322     & 1.182       & -1.091     & .2924      \\ \bottomrule
\end{tabular}%
 \begin{tablenotes}
\item [] \footnotesize \textit{Note}. Each variable was first standardized. We then performed a series of one-sample t-tests to test whether participants score on average above or below 0 (i.e., the average across all activities), separately for each activity and variable.
\end{tablenotes}
\end{threeparttable}
}
\end{table}

\begin{table}[]
\caption{Differences between activities (continued)}
\label{tab:b}
\resizebox{\textwidth}{!}{%
\begin{threeparttable}[t]
\sisetup{
    group-digits=true,
    detect-weight=true,
    detect-shape=true,
    table-format=-1.4,
    table-alignment = left
}
\begin{tabular}{@{}lSSSSSSSSSSSS@{}}
\toprule
               & \multicolumn{4}{l}{\textbf{Relatedness}}           & \multicolumn{4}{l}{\textbf{Pressure}}              & \multicolumn{4}{l}{\textbf{Conflict}}              \\
               & \textit{M} & \textit{SD} & \textit{t} & \textit{p} & \textit{M} & \textit{SD} & \textit{t} & \textit{p} & \textit{M} & \textit{SD} & \textit{t} & \textit{p} \\ \midrule
Coding         & 0.079      & 0.941       & 1.088      & .2783      & -0.138     & 0.958       & -1.997     & .0473      & 0.084      & 1.049       & 1.038      & .3008      \\
Bugfixing      & -0.077     & 0.998       & -0.748     & .4566      & -0.211     & 0.996       & -2.234     & .0275      & 0.009      & 0.937       & 0.093      & .926       \\
Testing        & -0.07      & 1.084       & -0.603     & .5479      & -0.035     & 1.074       & -0.323     & .7471      & -0.053     & 1.001       & -0.486     & .628       \\
Specification  & 0.332      & 0.942       & 1.577      & .1313      & -0.161     & 1.037       & -0.729     & .4739      & -0.205     & 0.96        & -1.001     & .3282      \\
Reviewing      & -0.024     & 0.995       & -0.216     & .8298      & 0.096      & 1.015       & 0.901      & .3702      & 0.163      & 1.06        & 1.385      & .1698      \\
Documenting    & 0.051      & 0.984       & 0.319      & .7513      & 0.11       & 0.904       & 0.77       & .4459      & -0.069     & 1.021       & -0.413     & .6817      \\
Meetings       & 0.217      & 0.861       & 2.39       & .0189      & -0.02      & 0.936       & -0.2       & .8418      & 0.1        & 1.014       & 0.936      & .3517      \\
Emails         & -0.527     & 1.074       & -3.292     & .002       & 0.261      & 0.977       & 1.91       & .0619      & -0.113     & 0.957       & -0.801     & .4271      \\
Helping        & 0.655      & 0.599       & 6.378      & 0          & 0.163      & 0.901       & 1.039      & .3066      & -0.053     & 0.976       & -0.312     & .7572      \\
Networking     & 0.38       & 0.812       & 1.479      & .1732      & -0.059     & 0.872       & -0.224     & .8273      & 0.059      & 1.022       & 0.173      & .8669      \\
Learning       & -0.298     & 1.028       & -2.476     & .0156      & 0.292      & 1.053       & 2.676      & .0088      & -0.233     & 0.921       & -2.181     & .0324      \\
Administration & -0.508     & 1.311       & -1.396     & .1879      & 0.022      & 1.145       & 0.07       & .945       & -0.161     & 0.94        & -0.593     & .5652      \\ \bottomrule
\end{tabular}%
 \begin{tablenotes}
\item [] \footnotesize \textit{Note}. Each variable was first standardized. We then performed a series of one-sample t-tests to test whether participants score on average above or below 0 (i.e., the average across all activities), separately for each activity and variable. 
\end{tablenotes}
\end{threeparttable}
}
\end{table}

\subsection{Exploratory Analysis}
We explored whether there are any gender mean differences for any of our activity-independent and activity-dependent variables, because other studies found that women's mental health and productivity were more negatively impacted by the Covid-19 pandemic than men's~\cite{carli2020women}. In total, we conducted 8 (activity-independent) + 201 (activity-dependent with $> 1$ women responding) independent samples t-tests. Because of the large number of comparisons, we adjusted our $\alpha-$threshold to .0005. None of the t-tests reached statistical significance, all $ps > .0006$. We report descriptive and relevant inferential statistics for each of the 209 t-tests in the Online Supplemental Materials on Zenodo.
Additionally, we explored whether day of the week is not only associated with productivity -- previous research found that productivity is higher Tuesdays to Thursdays and lower on Mondays and Fridays~\cite{senney2019role} -- but also associated with well-being or needs. However, this was not the case, according to a series of both Pearson's and Spearman's rank correlations $rs < .13, ps > .07$.

\section{Discussion}
\label{sec:discussion}

\subsection{Revised Theoretical Framework}

Our results partly align with the theoretical framework proposed by Deci et al.~\cite{deci2017self} (cf. Figure~\ref{fig:theory1}). Whereas the exploratory study did not find that activities are significantly correlated with needs or the dependent variables, the confirmatory study found support for it. We found that some activities were linked with the activity-specific needs of the self-determination theory as well as activity-specific satisfaction. Additionally, activity-specific needs were associated with activity-specific satisfaction and productivity. However, while our findings are in line with Deci et al.'s~\cite{deci2017self} broad framework, we are, to the best of our knowledge, the first in testing which activities show stronger links with activity-specific needs, satisfaction, and productivity.

\begin{figure*} 
\centering 
\includegraphics[width=1\textwidth]{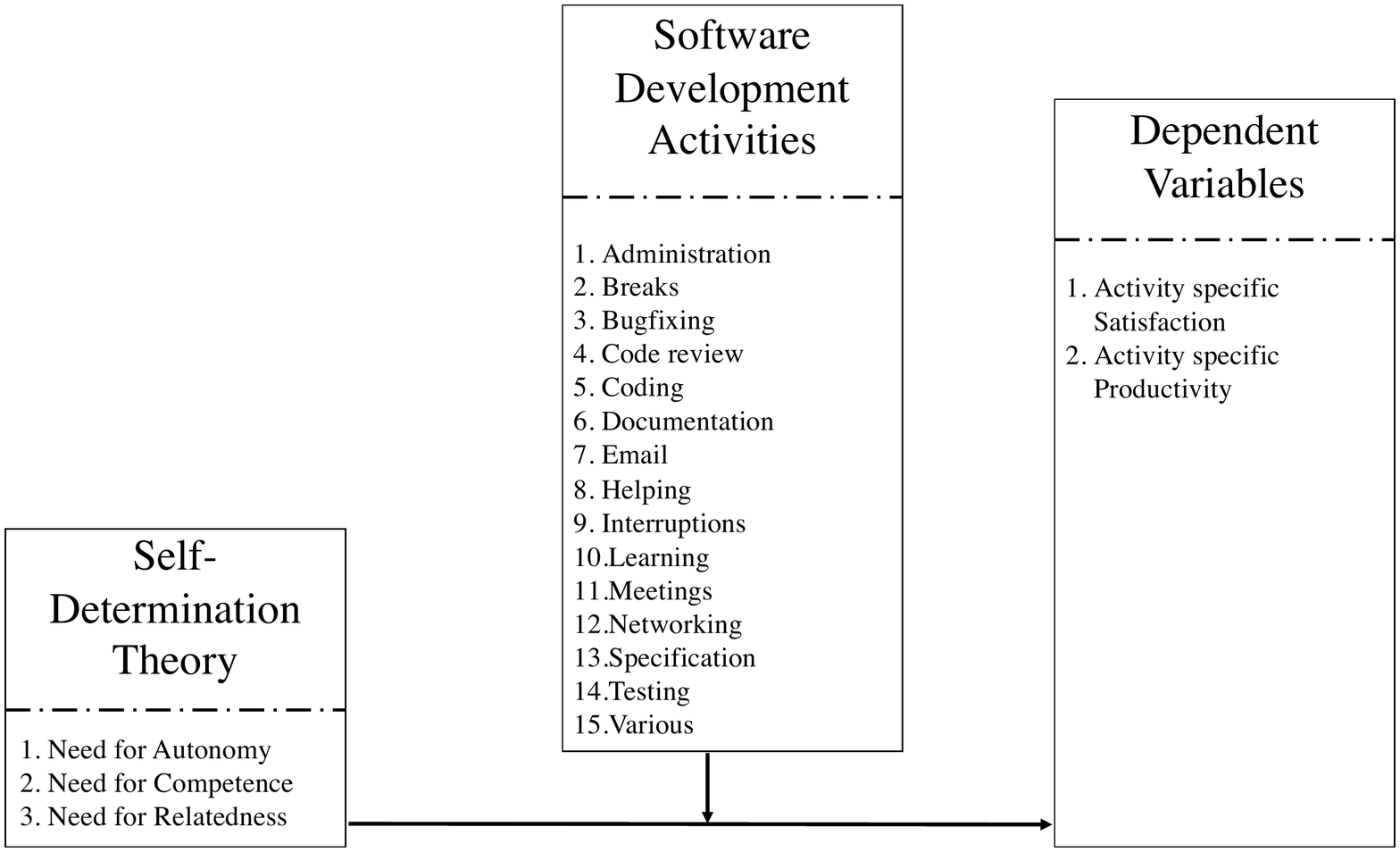} 
\caption{Revised Theoretical Framework. We found that the strength of the association between Self-Determination Theory needs and the two dependent variables depends on the type of activity performed by developers.} 
\label{fig:theory2} 
\end{figure*}

However, a revised theoretical framework is also supported by our confirmatory study: The links between the three needs and activity satisfaction as well as productivity are moderated by the type of activity (moderation is represented in Figure~\ref{fig:theory2}, a consequence of our findings of Figures~\ref{fig:coeff2} to~\ref{fig:coeff2-2}). In other words, the strength of the association between needs and activity-satisfaction as well as productivity depends on the type of activity.  
The model depicted in Figure~\ref{fig:theory2} does not directly contradict the model shown in Figure~\ref{fig:theory1}, but it revises it. They can co-exist, as our data shows. The model from Figure~\ref{fig:theory1} is more relevant to understand underlying mechanism and basic processes, whereas the model from Figure~\ref{fig:theory2} has more applied value. Indeed, the latter model offers intriguing possibilities for future research, which we discuss in more detail below.

\subsection{Implications for Research and Practice}

Our investigation addresses the need for scholarly evidence concerning the effects of WFH during the COVID-19 pandemic on software developers' work activities, including the impact on professionals' well-being and productivity.
Further, a deeper understanding of the effect of the pandemic on professional working life for the large number of software professionals working remotely provides relevant insights for both research and practice. 
To this end, this study makes several contributions, as summarized in Table~\ref{tab:Findings}.

\begin{center}
\small
\begin{tabularx}{\textwidth}{@{}XXX@{}}
\caption{Summary of key findings \& implications}
\label{tab:Findings} \\
    \toprule
   & \textbf{Findings} & \textbf{Implications} \\ \midrule
\endfirsthead
    \caption{Summary of key findings \& implications \hfill(continued)}            \\
    \toprule
   & \textbf{Findings} & \textbf{Implications} \\ \midrule
\endhead

  RQ1: Has the distribution of daily working activities of software engineers changed while WFH during the pandemic as compared to pre-pandemic daily working activities?
  & Overall, the ranking among work activities remains mostly unchanged. However, when WFH developers spend less time in: Bugfixing ($t_1=-5.31, t_2=-3.55$), Meetings ($t_1=-9.95, t_2=-6.63$), Breaks ($t_1=-7.39, t_2=-14.30$), Interruptions ($t_2=-5.39$), E-Mails ($t_1=-3.69$), and more time in Specification ($t_1=4.65, t_2=4.05$), ($t_1=4.65, t_2=4.05$), Testing ($t_1=3.41, t_2=3.32$), Administration ($t_1=4.58, t_2=4.28$), Documentation ($t_1=5.18, t_2=5.07$), Learning ($t_1=4.24, t_2=3.38$). Additionally, we found very high correlation of the group averages of time 1 and 2: $r(13) = .99, p < .0001$. A series of 15 paired t-tests comparing the relative time spend on each of the 15 activities between time 1 and 2 found little change. Two exceptions were more Breaks ($t=4.71$) and Networking ($t=4.33$) at time 1 compared to time 2. 
  & WFH does not affect the time spent on working activities by software developers, and the distribution is comparable to a typical office day. One interpretation might be that the significant time reduction of meetings suggests that online meetings are more time-efficient than physical ones. Also, professionals seem to be more focused when working remotely and have fewer interruptions. This allows them, among others, to dedicate more time to developing their own skills. Developers had a very regular work activity distribution during the pandemic, comparable to their office day. Fewer breaks and networking might suggest that professionals adapted to the new situations towards the end of the first lockdown in May 2020 in many countries, and became more time-efficient.   \\ 

  RQ2: Is the distribution of daily working activities related to well-being, productivity, and other variables? 
  & A series of $2\times 195$ correlation analyses did not show substantially significant results. Overall, we conclude that work activities carried out at home are not related to well-being, productivity, and other variables such as stress, boredom, or needs.
  & This can be interpreted as a generally positive finding, as it shows that various activities are unrelated to important psychological and social variables while WFH if they are measured typically (e.g., well-being over the past week).  \\  
    
  RQ3: Do the needs for autonomy, competence, and relatedness predict software engineers' activity-specific satisfaction and productivity?
  & In the confirmatory study, we found, across all activities, that the needs for autonomy, competence, and relatedness were positively associated with activity satisfaction and productivity, using linear mixed-effects modeling and multiple linear regression analysis. Conflict and pressure were only negatively associated with activity-specific satisfaction but unrelated with activity-specific productivity. These associations were primarily consistent across activities, albeit a few deviations occurred (Fig.~\ref{fig:coeff2} and~\ref{fig:coeff2-2}). 
  & Self-determination theory provides a robust framework to understand and enhance developers' productivity and well-being.
    A higher degree of autonomy, competence, and relatedness for software professionals can increase their satisfaction and productivity.
    Rather than control or micro-management, organizations should support employees to tailor their own working activities and training.  \\ 

  RQ4: Are the associations between activity satisfaction and productivity moderated by resilience and company support?
  & A series of 80 moderated regression analyses revealed that neither caring leadership, work-life balance, empowerment, job enablement, soft company support, hard company support, nor recognition moderates the link between the three needs and activity satisfaction and productivity. Additionally, all seven task-unrelated variables were unrelated to activity-specific satisfaction and productivity.
  & Our results are inconclusive. Possibly, with more specific measures (e.g., activity-specific company support), this outcome might change. As a community, we need better and more nuanced measurements of satisfaction and productivity to identify specific factors that contribute to professionals' satisfaction and productivity compared to overall assessments. Repeated self-reports (\textit{e.g.} or Experience Sampling~\cite{larson2014experience}) can identify the effect of contextual factors (\textit{e.g.} current task). This allows for collecting reliable and contextually rich data as participants assess their current state rather than reflect on an extensive time in the past~\cite{Berkel2020HumanAccuracy}.   \\

  RQ5: Do software engineers' work activities while WFH during the pandemic affect their activity-specific well-being, productivity, and psychological needs?
  & We found that activity satisfaction was relatively lower when participants were bugfixing and higher when they were helping others, using a series of 84 one-sample t-tests. Additionally, autonomy was perceived lower while professionals were in meetings or writing emails. Competence was higher when professionals were helping others and lower when bugfixing. Relatedness was higher when professionals were helping others. The findings hold even after controlling for multiple comparisons.
  & Bugfixing is associated with lower activity satisfaction while helping improves it. Code review, innersourcing, mentoring projects, and bug triaging processes support software engineers' desire to help, making them more satisfied and productive. At the same time, more junior figures can learn from more experienced ones, increasing employees' retention, and the helpers' satisfaction. \\ \bottomrule
\end{tabularx}
\end{center}

First, we ran an exploratory longitudinal study during the COVID-19 lockdown with 192 carefully selected software professionals to address the first and second research questions.
We assessed developers' working activities and their perceived well-being, productivity, and other relevant psychological and social variables.
Our data quality was assured by the high test-retest reliability of each variable, measuring at least $.50$, and Cronbach's alpha values above $.60$.

Second, we compared the time spent on typical office-based activities with the same activities while working from home.
Using the taxonomy and previously collected data of Meyer et al.~\cite{meyer2019today}, we ran 30 one-sample $t$-tests to assess significant differences.
Although we reported several differences, they are relatively small, which indicates that the time spent on different activities is almost identical in both the online and the physical working environment.

Third, we analyzed whether the time spent on each working activity changed during the pandemic.
After performing 15 paired $t$-tests, we conclude that developers did not change how they spend their time over a period of two weeks.

Fourth, we investigated whether well-being-related variables and productivity are associated with the time spent on each activity and if the findings replicate across both time points.
To do so, we ran twice 195 correlation analyses.
Our results suggest that well-being-related variables and productivity are not associated with the time spent on each activity. 

However, a shortcoming of our exploratory study is that we only measured general well-being, productivity, and needs and the amount of time spent on various activities during the past week. The lack of significant findings could suggest that either the type of activity does not impact professionals' well-being and productivity or that many other factors impact well-being and productivity more strongly (e.g., quality of social contacts~\cite{miller2021your,russo2020predictors}). We found evidence for the former in our confirmatory study.  

In our confirmatory study, we tested whether activity-specific variables, such as the need for autonomy, competence, relatedness, and activity-independent variables, such as resilience or empowerment, are associated with activity satisfaction and productivity (to address the third research question). Additionally, we tested whether activity-specific and activity-independent variables interact in predicting activity satisfaction and productivity, addressing the fourth research question. Finally, we tested whether specific activities impact professionals' activity-specific satisfaction and productivity, addressing the fifth and final research question.
Here, we summarize and discuss the results of our research questions.

\textit{\textbf{RQ1}: Has the distribution of daily working activities of software engineers changed while WFH during the pandemic as compared to pre-pandemic daily working activities?}

On the whole, we did not register significant changes to developers' work distribution. 
Further, we highlight that Meyer et al.'s sample covers only one software company (Microsoft)~\cite{meyer2019today}, whereas we surveyed developers across many companies globally distributed.
Therefore, some deviations were expected.
Nevertheless, we still report an overall consistency between our WFH data and Meyers et al.'s analysis of a typical office day at Microsoft.
Our results show that neither working from home nor sample type (Micosoft employees vs a diverse sample) affected how software engineers dedicate their time to specific activities.  
However, we observed some minor differences. Most notably, software engineers in our sample spend less time on bugfixing, meetings, and breaks.
Also, they report less time on e-mail writing (only in wave 1) and fewer interruptions when working from home (only in wave 2).
In contrast, they spend more time on specifications, testing, administration, documentation, and learning. It is unclear whether those minor differences emerged because of the pandemic or because our sample differed. 

We observe that meetings are significantly reduced while working remotely.
One explanation is that they are, on average, shorter and more time-efficient than in the office. For example, small talk might be perceived as more challenging during online meetings than in-person meetings. Alternatively, they might be better planned since setting up online meetings often requires a clear start and end time. 
Also, our participants invested in improving their skill set as they spent more time learning.
Similarly, developers seem to be more focused on their activities: They reported fewer breaks and interruptions.
At the same time, developers remain linked to their organization or their colleagues since their time on networking remains the same.
We did not register any significant change in the work activities during our exploratory investigation, with only two exceptions: at the first wave, developers spent more time on breaks and networking than during the second wave. 
Nevertheless, we report a correlation close to 1 of the group averages, suggesting a very high consistency in the pandemic activity distribution.
The reason software engineers spent less time on breaks and networking during the second measurement point might indicate that they became more accustomed to their new WFH condition.
Accordingly, professionals learned to spend their working time more efficiently.
Similar conclusions are also supported by the literature~\cite{ford2020tale,russo2020predictors}.

\textit{\textbf{RQ2}: Is the distribution of daily working activities related to well-being, productivity, and other variables?}

We did not find any significant relations between daily working activities and the well-being, productivity, or other investigated variables, except for one, taking breaks was negatively associated with productivity in our exploratory study. 
This can be interpreted as a generally positive finding since it shows software engineers' well-being and productivity do not depend on the type of activity they are doing, at least concerning the 15 measured activities.
The only significant relation was productivity, which correlated negatively with breaks in wave 1.
Despite being intuitive, we are very cautious about concluding that developers should take fewer breaks to be more productive since such a relation was not significant at wave 2 (although still negative). Further, prior work shows that breaks can increase well-being~\cite{dababneh2001impact} and can improve the quality of professionals' social networks~\cite{waber2010productivity}. Similarly, correlation does not equate to causation: participants might have taken more breaks because they felt less productive for various reasons (e.g., more exhaustion, distractions at home). 

Regarding the other activities, we do not find evidence that the time spent on activity affects productivity or well-being. 
We did not register any significant effect on how the amount of time dedicated to development activities impacts software engineers' general well-being, stress, boredom, or distractions while working from home. 
Previous studies showed that during the pandemic, it is essential to have daily routines to improve personal well-being~\cite{russo2020predictors}.
However, routines do not seem to play a significant role when it comes to individual activities.
As our findings show, possible distractions that might happen while working from home (e.g., children, noisy neighbors) do not influence the time spent on specific work activities.

The innate psychological needs of the self-determination theory~\cite{ryan2000self}, and its three dimensions, need for autonomy, competence, and relatedness, are associated with work motivation in general~\cite{gagne2005self}. 
To the best of our knowledge, our study is the first in our community to assess whether specific activities are correlated with autonomy, competence, and relatedness. 
Overall, we found that general psychological needs were unrelated to the amount of time developers spent on specific activities. 
In hindsight, this might be because the scale we used to measure the three dimensions of the self-determination theory captures broad human needs in general~\cite{ryan2000self} and not specifically while working on specific activities. We addressed this limitation of the exploratory study in the confirmatory analysis. 

While working remotely, the quality of communication between team members can be challenging, as face-to-face communication has to pass through a medium (e.g., MS Teams, Zoom).
Therefore, not being directly connected to the organizations can become a big issue for remote workers.
For example, research suggests that lower support from coworkers and supervisors~\cite{mccalister2006hardiness}, perceiving the values of one's organization to be different from one's values~\cite{edwards2009value}, and unfair treatment and lack of appreciation~\cite{bhui2016perceptions} are putting the mental health of remote workers at risk. 
Interestingly, our results suggest that the quality of communication does not relate to individual working activities.
This might seem surprising at first glance, as it is plausible to assume that those who find the quality of communication to be poorer might engage less in activities that require more communication (e.g., meetings) and more in activities that require less direct communication (e.g., coding, bugfixing). 
This might suggest that developers are professional enough not to let their behavior be influenced by their perception of the quality of communication. In other words, the time spent by software engineers on each activity is not detrimental to the relations with their organization.
Prior research has mostly ignored whether activity type plays a role in professionals' psychological and social factors. 
Typically, scholars only measured whether people are, for example, overall stressed, as opposed to stressed by specific activities~\cite{bhui2016perceptions,edwards2009value,mccalister2006hardiness}. 
Our research suggests that the type of activity is not a confounding variable, which increases our trust in prior research, which has typically looked at subjective work experience in general rather than actual activities. 
So, our exploratory findings suggest that software engineers' psychological and social factors do not matter on \textit{what} work activity they are performing, but rather \textit{how} it is done.


\textit{\textbf{RQ3}: Do the needs for autonomy, competence, and relatedness predict software engineers' activity-specific satisfaction and productivity?}

In the confirmatory study, we found, across all activities, that the needs for autonomy, competence, and relatedness were positively associated with activity satisfaction and productivity. Simultaneously, conflict and pressure were only negatively associated with activity satisfaction but unrelated to productivity. These associations were mostly consistent across activities, albeit a few deviations occurred (Fig.~\ref{fig:coeff2} and~\ref{fig:coeff2-2}). For example, relatedness predicted activity productivity for meetings and reviewing but not for coding, bug fixing, testing, and learning. One possibility is that meetings and reviewing are typically more social (i.e., done with other people), making relatedness more relevant.
Overall, our results align with our first findings, even though previous research often used different measures of well-being and/or needs, a variety of statistical tests (e.g., zero-order correlations), and/or relied on different populations such as student samples. For example, a meta-analysis~\cite{yu2018general} found a correlation of $r = .49, 95\%-CI[.39, .57]$, between need for autonomy and satisfaction with life, which is very much in line with our findings: A linear mixed-effects model with only autonomy as a predictor for activity satisfaction revealed $\beta = .48$\footnote{$\beta$ is standardized effect size, as is the correlation coefficient $r$.}.

This result is of great relevance to understanding developers' satisfaction and productivity.
To improve activity satisfaction and productivity, self-determination theory is a precious lens.
Indeed, more autonomous, competent, and related professionals show a high degree of satisfaction and productivity.
These findings are also precious for employee recruitment and retention. 
Companies should keep this aspect in mind when organizing working activities.
In particular, micro-management could be detrimental to software engineers' satisfaction and productivity.
In other words, it is advisable to discuss realistic working goals of software projects, leaving it to the teams to self organize, like a recent investigation about effective Scrum teams highlighted~\cite{verwijs2021theory}.

\textit{\textbf{RQ4}: Are the associations between activity satisfaction and productivity moderated by resilience and company support?}

None of the seven task-unrelated variables (e.g., resilience, work-life balance) moderate the link between the three needs and activity satisfaction and productivity. Initially, we hypothesized that, for example, resilience might buffer against reduced autonomy because resilient people are more likely to bounce back after stressful events such as being less able to make autonomous decisions~\cite{smith2008brief,weinstein2011self}. 
Generally measured variables (e.g., general work-life balance) are rarely associated with specific variables~\cite{davidson1979variables}. This might be because we measured resilience and work-life balance in a way that is too broad. Future research could measure resilience in a more specific way (e.g., resilience during the day or activity-specific resilience), which makes it more relevant for activity-specific satisfaction, productivity, and basic needs. 
Alternatively, other personality traits might be more relevant. For example, proactive personality was found to mitigate or moderate the effect between stress and productivity~\cite{hung2015does,onyemah2008role}. Thus, lower levels of activity satisfaction might strongly impact productivity for those who score low on proactive personality. \par
Additionally, caring leadership, work-life balance, empowerment, job enablement, soft company support, hard company support, and recognition were unrelated to activity-specific satisfaction and productivity. In hindsight, this is not surprising given that we measured all these variables generally. For example, if we had measured activity-life balance instead of general work-life balance, we would have likely found an effect on activity-specific satisfaction and productivity.

Overall, our results are inconclusive on this question. 
Although the moderation effects of resilience and company support are not supported, we acknowledge that with more specific measurements, this outcome might change.

\textit{\textbf{RQ5}: Do software engineers' work activities while WFH during the pandemic affect their activity-specific well-being, productivity, and psychological needs?}

We found that activity satisfaction was relatively lower when participants were bugfixing and higher when helping others. This finding is in line with previous research suggesting that helping others increases well-being~\cite{buchanan2010acts}. In contrast, levels of activity productivity were more consistent across activities, while activity satisfaction varied. 
Our findings of bugfixing have three main practical implications. 

First, bugfixing might be viewed as an annoying but necessary activity by many developers: Compared to all activities, 80 participants reported a below-average level of satisfaction when bugfixing, whereas only 35 reported above-average satisfaction (cf. Fig.~\ref{fig:tasks1}). 
Pointing out the meaningfulness of bugfixing is essential. Literature supports that meaning is positively associated with satisfaction, autonomy, competence, and relatedness~\cite{martela2018meaningfulness}. Even though most developers are aware that bugfixing is essential, the odd reminder or nudge can have an impact~\cite{Venema2021}. For example, while most people are aware that switching off the light when needed is beneficial for the environment, reminders of it nevertheless increase the likelihood that the light gets switched off~\cite{byerly2018nudging}. Occasional reminders or nudges are typically very inexpensive and are likely to be cost-effective. However, more research is needed whether in the context of bugfixing nudges result in substantially increased satisfaction.
Additionally, organizations should support a higher degree of socialization during bugfixing activities.
Software engineers appear to be (contrary to stereotypes) social and caring individuals. 
Consequently, code review practices should be primarily supported by management.

Second, organizations should facilitate an inclusive working environment in which developers are actively helping each other to perform different activities they can freely choose from.
One concrete example might be to establish innersourcing projects~\cite{stol2014inner}. They are similar to open source projects, except they are closed projects in which only employees can participate.
This practice would also support the need for autonomy of software professionals in contributing to projects they find important and committed to.

Third, establishing mentorship programs can stimulate senior developers' desire to help by increasing newcomers' sense of relatedness.
This aspect is even more critical in a WFH setting, where informal networking occasions are typically limited.
At the same time, this will increase the onboarding success of new employees.
Research already showed that the support of newly hired employees through, for example, mentoring projects, is an essential factor for onboarding success and, eventually, employees' retention~\cite{sharma2020exploring}.
Furthermore, an effective bug triaging process is considered pivotal for a software organization efficacy to address quality concerns~\cite{anvik2006should}. 
Picking the right developer to work on a specific bug is crucial to fix the bug timely and to reduce bug tossing length~\cite{yadav2019ranking}.
Establishing an effective and transparent process is, thus, a way to establish meaning about this activity. 
Future research along the lines of RQ5 could also investigate whether an activity was self-chosen. If an activity is self-chosen, intrinsic motivation is usually higher, which is linked to higher job satisfaction and performance~\cite{hayati2012islamic}.

\subsection{Measuring satisfaction and productivity}
Findings from both studies have not only practical but also methodological implications. 
The time developers spend on a specific activity was unrelated to their well-being, productivity, needs, or working conditions, when the latter was measured in a general (i.e., activity unrelated) way. Researchers or employers who wish to identify how to increase satisfaction or productivity of a specific activity need to adapt their measures to become activity-specific. For example, increasing employees' general resilience or work-life balance will have little impact on how satisfied and productive they are with a specific coding task. In contrast, enhancing autonomy for coding is likely more beneficial. However, it should be noted that we have created the activity-related measures for the confirmatory study. While the measures were mainly associated with each other in the expected directions, even when measured with a single item (e.g., conflicts and pressure were negatively associated with activity satisfaction), future research could further improve our measures to increase their reliability.  \par
However, this does not imply that general measures of personality and other constructs cannot predict activity-specific variables. 
Previous research established that, for example, personality variables predict related behavior averaged over a sample of occasions and situations much better than single observations~\cite{epstein1979stability,skimina2019behavioral}. In contrast, averaging across multiple instances of autonomy-related behaviors across various situations (e.g., living in a self-chosen city, working in an area that matches personal interest, or listening to the music one likes most as opposed friends, partner, or family) will likely be more strongly associated with need for autonomy. For example, looking at an exhibition that is of personal interest at a museum might only weakly be predicted by need for autonomy.  
This is because general measures are broad and trans-situational by definition. 
For instance, resilience is important in many aspects of a software developer's life, not only while coding on a specific day. This activity, in turn, can also be influenced by many situational variables (e.g., distractions at home, a particular project, working with competent colleagues) that diminish the impact of personality. If researchers are interested in testing whether, for example, resilience predicts activity satisfaction, they might want to measure activity satisfaction across multiple activities (e.g., coding, bugfixing) and/or multiple time-points~\cite{Berkel2019ContextSelf}.

Further, our findings cautiously suggest that WFH might be more beneficial for both developers and organizations than working in the office, or at least for some groups of professionals~\cite{ford2019remote}. However, while some studies support our conclusion that WFH increases or does not impact productivity~\cite{bao2020does,barrero2021working,deole2021home,russo2020predictors}, some studies also found that WFH has a negative impact on productivity~\cite{gibbs2021work,kitagawa2021working,morikawa2020productivity}. As there are too many potential differences between the studies (e.g., cultural factors, working conditions at home, type of work, measurement of productivity), cross-country and cross-profession studies are needed.
Large sample sizes or meta-analyses synthesize the findings to better understand the conflicting findings in the direction WFH shifts productivity during the Covid-19 pandemic. Thus, there is a need for more research to identify factors that help us understand how WFH can be beneficial and whether these factors are transferable to working on-site. Our confirmatory study offers an intriguing possibility for the contradictory studies: The type of activity matters. Certain activities might be less feasible when WFH, which reduces productivity, whereas working on other activities might be more accessible when WFH and thus increase productivity, similar to what we predict in Figure~\ref{fig:theory2}.

\subsection{Threats to validity}
To conclude this section, we briefly address the most relevant limitations. 

\textit{Reliability}. We investigated our subject matter using a longitudinal exploratory design combined with a confirmatory cross-sectional one.
Participants were identified using a multi-stage selection process to ensure (i) they are professionally active software engineers, (ii) data quality, and (iii) that they were working from home during the lockdown.
Validated scales have been used when available or adapted from previous investigations.
In line with most related research, we have not aimed to control for response biases because doing so has usually little impact on the reliability: Some approaches to control for response bias improve the reliability slightly, but can also reduce reliability or leave it unchanged~\cite{he2017enhancing}.
Overall, we report a high test-retest reliability in the longitudinal study and adequate internal consistencies of all measures.

\textit{Construct validity}. 
To enhance cross-study comparability, we used the taxonomy by Meyer et al.~\cite{meyer2019today} to define the daily activities of software developers.
Similarly, we used those benchmarks to confront it with working from home settings.
However, we did not monitor developers' effectiveness by executing every activity while working remotely.
We opted for this to be consistent with Meyer et al. and because we collected data from a global sample of software professionals working in 190+ different organizations, making the development of objectively comparable measurements near impossible.
Still, we report some differences with the data collected by Meyer et al., although the difference is of only some percentage points. 

\textit{Conclusion validity}.
Our conclusions rely on multiple statistical analyses, such as one-sample $t$-tests, paired $t$-tests, Pearson's correlation, multiple regressions, and linear mixed-effects models.
Furthermore, we also ran a non-parametric Spearman's rank correlations test for our conclusion's consistency since not all distributions were perfectly normally distributed.
To support Open Science, we make a reproducible R-code alongside our raw data openly available on Zenodo.

\textit{Internal validity}.
We used self-reported measures for well-being, productivity, and other psychological and social variables for this investigation, which might be considered a limitation.
The data for the exploratory study was collected towards the end of the first lockdown in spring 2020 with a longitudinal design.
We expanded our initial data collection one year later, in spring 2021, with a cross-sectional, confirmatory study.
This enabled our participants to report a more mature and stable assessment of the new working setting.
For the exploratory investigation, we only considered countries with comparable lockdown measures (e.g., we excluded, among others, Denmark, Germany, and Sweden as these countries did not face a total lockdown or had different measures in place in the country's regions).
Thus, we asked both waves about lockdown conditions in their home country and if they were still working from home.
Moreover, the exploratory longitudinal study was performed in a relatively short time frame (around two weeks) due to the ever changing health public policies in the first months of the pandemic. 
We do not deem it as a significant limitation, since the main goal of this first study was to identify relevant tendencies to follow up in the confirmatory study.
Since all selected informants faced comparable conditions, we did not exclude any of the 192 selected software professionals.
For the confirmatory study, we surveyed 300 developers working from home.
Since lockdown measures in spring 2021 were comparable across all countries, we did not exclude any country \textit{a priori}.

\textit{External validity}.
We designed this study to maximize internal validity.
Therefore, we determined our sample size with an \textit{a priori} power analysis.
So, we did not work with a representative sample of the software engineering population in mind (such as Russo and Stol~\cite{russo2020gender} did, where the research goal was to generalize results, surveying over 400 software engineers).
However, we recognize having submitted our surveys in the middle of a very peculiar period. 
This makes it unclear whether we can generalize our findings to non-pandemic working from home settings.
Notwithstanding, we also realize that we require fast and reliable evidence regarding the COVID-19 crisis we are facing right now, improving the quality of developers' daily lives.
This study will also enable a better-informed research design for future remote working studies once this pandemic is over.
Finally, our sample is almost entirely composed of western country developers. 
Consequently, the investigated effects could be different in other regions of the world e.g., Africa or Asia.

\section{Conclusion}
\label{sec:conclusion}

This research focused on software engineers' activity satisfaction and performance during the COVID-19 pandemic.
For the sake of clarity, we did not provide any consideration regarding the Future of Work after the COVID-19 lockdowns, such as Smite et al.~\cite{smite2023future}.
To do so, we first employed an exploratory longitudinal study design across two waves and a confirmatory cross-sectional study. We found that developers still spend proportionally the same amount of time on their different daily activities. 
For example, the software engineers in our sample still spent most of their working time on coding, bugfixing, meetings, testing, and e-mails, as previously reported by Meyer et al.~\cite{meyer2019today}.
Nevertheless, we found some significant mean differences. Our participants reported having spent less time in meetings and breaks, suggesting that both were less common, possibly due to developers' adaption of working remotely.
Similarly, no significant relations have been found between productivity, well-being, and relevant social and psychological variables with working activities.
In our confirmatory cross-sectional study, we found that activity-specific needs for autonomy, competence, and relatedness are associated with activity-specific satisfaction and productivity. Furthermore, activity satisfaction was relatively lower when participants were bugfixing and higher when helping others. At the same time, autonomy was perceived as relatively lower while professionals were in meetings or writing e-mails. 

Overall, our research suggests that WFH does not \textit{per se} affect how much time developers spend working on various activities.
Nevertheless, software engineers are social beings, and their satisfaction increase when they can help others.
This paper also suggests a number of recommendations for organizations to support their employees' well-being and productivity.
In particular, active company policies to support developers' need for autonomy, relatedness, and competence appear to be particularly effective in a WFH context.
Also, bugfixing is the most detrimental activity for professionals' satisfaction.
Accordingly, specific processes should be designed for software engineers working from home (e.g., bug triaging and mentorship programs).

As a deductive investigation, using a quantitative stance we can only assess the relations between the independent variables with our two dependent ones. 
Thus, we do miss a number of nuances about the interactions of our variables that should be investigated further. 
Additionally, future research should aim to provide more tailored recommendations based on developers' personalities. This would result in a more nuanced understanding of the subject matter.
Also, a better understanding of software professionals' activity satisfaction and productivity is needed to develop reliable measurement instruments and to develop or refine theories.

\section*{Supplementary Materials} 
The complete replication package is openly available under CC BY 4.0 license on Zenodo, DOI: \url{https://doi.org/10.5281/zenodo.7298506}.


\section*{Acknowledgment}
This work was supported by the Carlsberg Foundation under grant agreement number CF20-0322 (PanTra --- Pandemic Transformation).

\section*{Conflict of Interest}
All authors declare that they have no conflicts of interest.

\bibliographystyle{spmpsci}   
\bibliography{main}  

 \clearpage 
 \appendix
 \section{Appendix}
 \begin{figure}[h]
    \centering
    \includegraphics[width=1\textwidth]{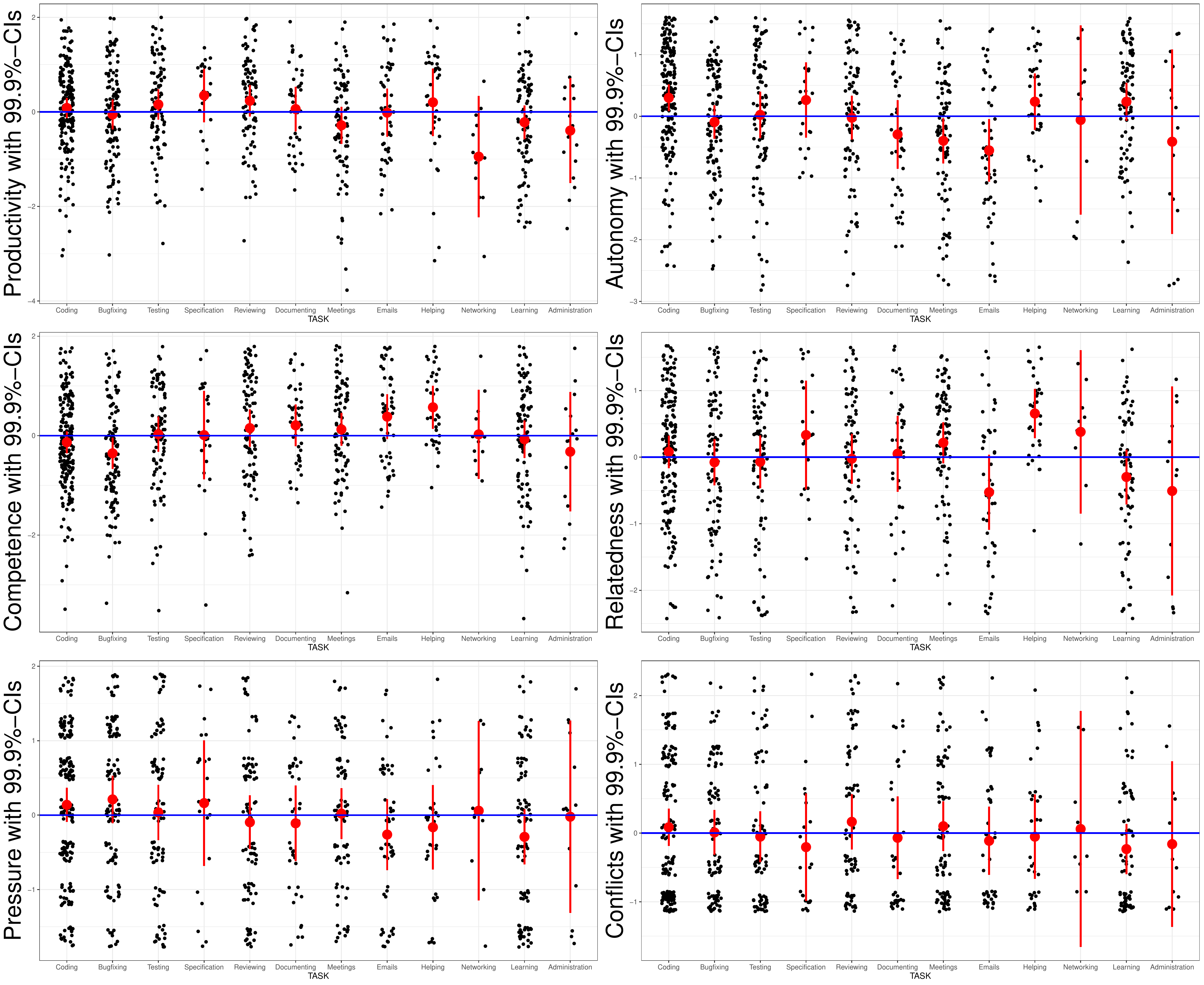}
    \caption{Differences between activities regarding productivity, autonomy, competence, relatedness, pressure, and conflicts. Red lines represent 99.9\%-CIs.}
    \label{fig:tasks2}
\end{figure}

\end{document}